\documentclass[preprint,prd,aps,showpacs,showkeys,nofootinbib]{revtex4}
\usepackage{graphicx}
\begin{document}

\title{GKZ hypergeometric systems of the four-loop vacuum Feynman integrals}

\author{
Hai-Bin Zhang$^{1,2,3,4,}$\footnote{hbzhang@hbu.edu.cn},
Tai-Fu Feng$^{1,2,3,5,}$\footnote{fengtf@hbu.edu.cn}
}

\affiliation{
$^1$Department of Physics, Hebei University, Baoding, 071002, China\\
$^2$Hebei Key Laboratory of High-precision Computation and Application of
Quantum Field Theory, Baoding, 071002, China\\
$^3$Hebei Research Center of the Basic Discipline for Computational Physics, Baoding, 071002, China\\
$^4$Institute of Life Science and Green Development, Hebei University, Baoding, 071002, China\\
$^5$Department of Physics, Chongqing University, Chongqing, 401331, China
}

\begin{abstract}
Basing on Mellin-Barnes representations and Miller's transformation, we present the Gel'fand-Kapranov-Zelevinsky (GKZ) hypergeometric systems of 4-loop vacuum Feynman integrals with arbitrary masses. Through the GKZ hypergeometric systems, the analytical hypergeometric solutions of 4-loop vacuum Feynman integrals with arbitrary masses can be obtained in neighborhoods of origin including infinity. The analytical expressions of Feynman integrals can be formulated as a linear combination of the fundamental solution systems in certain convergent region, which the combination coefficients can be determined by
the integral at some regular singularities, the Mellin-Barnes representation of the integral, or some mathematical methods.
\end{abstract}


\maketitle

\newpage

\tableofcontents

\newpage

\section{Introduction\label{sec1}}

With the improvement of experimental measurement accuracy at the planned future colliders \cite{CLIC,ILC,CEPC,FCC,HL-LHC,Heinrich2021}, Feynman integrals need to be calculated beyond two-loop order. Vacuum integrals are the important subsets of Feynman integrals, which constitute a main building block in asymptotic expansions of Feynman integrals \cite{Smirnov2002,Misiak1995}, and may represent boundary conditions within the application of the differential equation method. Generically, vacuum integrals are important both for investigating formal properties of quantum field theories and for computational aspects within the evaluation of scattering, such as the structure of phase factors of the S-matrix operator and the UV behaviour of scattering amplitudes. The calculation of multi-loop vacuum integrals is a good breakthrough window in the calculation of multi-loop Feynman integrals. In this article, we investigate the analytical calculation of 4-loop vacuum integrals with arbitrary masses.

It's well known that the completely one-loop integrals are analytically in the time-space dimension $D=4-2\varepsilon$ \cite{tHooft1979,Passarino1979,Denner1993,V.A.Smirnov2012}.
At the two-loop level, the vacuum integrals have been calculated to polylogarithms or
equivalent functions \cite{R2loop1,R2loop2,R2loop3,R2loop4,R2loop5}. The three-loop vacuum integrals are also calculated analytically and numerically in some literatures
\cite{R3loop1,R3loop2,R3loop3,R3loop4,R3loop5,R3loop6,R3loop7,R3loop8,R3loop9,R3loop10,R3loop11,R3loop12,
R3loop13,R3loop14,R3loop15,R3loop16,R3loopN1,R3loopN2,R3loopN3,Gu2019,Gu2020,Zhang2023}.
But, very few four-loop vacuum integrals are calculated analytically. The vacuum integrals at the four-loop level are only calculated analytically for single-mass-scale \cite{R3loop10,R4loop2}, equal masses \cite{R4loop3}, and reduction \cite{R4loop4,R4loop5}.
Recently, Feynman integrals using auxiliary mass flow numerical method \cite{AMFlow1,AMFlow2}, also can be reduced to vacuum integrals, which can be numerical solved by further reduction. In order to improve the computational efficiency and give analytical results completely, it is meaningful to explore new analytical calculating method of the multi-loop vacuum integrals with arbitrary masses.

During the past decades, Feynman integrals have been considered as the generalized hypergeometric functions~\cite{Regge1967,Davydychev1,Davydychev3,Davydychev1991JMP,Davydychev1992JPA,Davydychev1992JMP,
Davydychev1993,Berends1994,Smirnov1999,Tausk1999,Davydychev2000,Tarasov2000,Tarasov2003,Davydychev2006,Kalmykov2009,
Kalmykov2011,Kalmykov2012,Bytev2015,Bytev2016,Kalmykov2017,Feng2018,Feng2019,Abreu2020,
Ananthanarayan2020,Ananthanarayan2021}.
Considering Feynman integrals as the generalized hypergeometric functions,
one can find that the $D-$module of a Feynman integral~\cite{Kalmykov2012,Nasrollahpoursamami2016} is isomorphic to Gel'fand-Kapranov-Zelevinsky (GKZ) $D-$module~\cite{Gelfand1987,Gelfand1988,Gelfand1988a,Gelfand1989,Gelfand1990}.
GKZ hypergeometric systems of some Feynman integrals
are presented in Refs.~\cite{Cruz2019,Klausen2019}
through Lee-Pomeransky parametric representations~\cite{Lee2013}.
To construct series solutions with suitable independent variables,
one can compute the restricted $D$-module of GKZ-hypergeometric system originating from
Lee-Pomeransky representations on corresponding hyperplane in the parameter
space~\cite{Oaku1997,Walther1999,Oaku2001}. But the constructed hypergeometric series solutions through Lee-Pomeransky polynomials may be non-canonical series, such as two-loop sunset diagram.
In our previous work, from Mellin-Barnes representations~\cite{Feng2018,Feng2019}, GKZ hypergeometric systems of one- and two-loop Feynman diagrams
also can be obtained~\cite{Feng2020,GKZ-2loop,Grassmannians}, through Miller's transformation~\cite{Miller68,Miller72}. Through GKZ hypergeometric systems of Mellin-Barnes representations, we can obtain the
fundamental solution systems consisting of canonical hypergeometric functions. There are some recent work in the GKZ framework of Feynman integrals
\cite{Klemm2020,Bonisch2021,Borinsky2020,Kalmykov2021,Tellander2021,Klausen2021,Mizera2021,
Chestnov2022,Chestnov2023,Walther2022,Munch2022,Ananthanarayan2022GKZ,
Klausen2023,Caloro2023,Munch2024}.

In our previous work, we have given GKZ hypergeometric systems of the Feynman integrals of the two-loop vacuum integral \cite{Feng2020} and three-loop vacuum integrals \cite{Zhang2023}. In this article, we further derive GKZ hypergeometric systems of the four-loop vacuum integrals with arbitrary masses, which will be helpful for further research in higher order analytic calculation for Feynman integrals with arbitrary masses.

Our presentation is organized as following. In Sec. \ref{sec-strategy}, we describe the computational strategy for GKZ hypergeometric system of Feynman integral, where the general algorithm and techniques used in the various steps of the algorithm are introduced. Then through the Mellin-Barnes representation and Miller's transformation,
we derive the GKZ hypergeometric system of the
four-loop vacuum integrals with five propagates in Sec. \ref{sec-five}, six propagates in Sec. \ref{sec-six}, and seven propagates in Sec. \ref{sec-seven}.
And then, we construct the hypergeometric series solutions of the GKZ hypergeometric systems of the four-loop vacuum integrals in Sec. \ref{sec-solu}.
In Sec. \ref{sec-con}, the conclusions are summarized. Some formulates
are presented in the appendices.

\section{Computational strategy for GKZ hypergeometric system of Feynman integral\label{sec-strategy}}

In this section, we firstly show the computational strategy for GKZ hypergeometric system of Feynman integral. The expression of $l$-loop Feynman integral with $N$-propagates can be written as
\begin{eqnarray}
U_{_N}=\Big(\Lambda_{_{\rm RE}}^2\Big)^{(2-D/2)l}\int{d^D{\bf q}\over(2\pi)^D}
{1\over D_{_1}^{\nu_{_1}}D_{_2}^{\nu_{_2}}\cdots D_{_N}^{\nu_{_N}}}\:,
\label{FI-1}
\end{eqnarray}
where the numerator is 1, $\Lambda_{_{\rm RE}}$ denotes the renormalization energy scale and $D=4-2\varepsilon$ is the number of dimensions in dimensional regularization, the loop momentum ${\bf q}=(q_{_1},\;\cdots,\;q_{_l})$,  ${d^D{\bf q}\over(2\pi)^D}=\prod\limits_{i=1}^l {d^Dq_{_i}\over(2\pi)^D}$. Here $D_{_i}=k_{_i}^2-m_{_i}^2$, where $m_{_i}$ denotes the virtual mass of the $i$-th internal particle and $k_{_i}$ denotes a linear combination
of the loop momentums and external momentums.

The multi-loop Feynman integral is hard to calculate analytically, if all virtual masses are nonzero. So, one can extract the virtual masses from the integral, to facilitate further calculation. One effective method is Mellin-Barnes transformation
\begin{eqnarray}
&&{1 \over (A+B)^{\nu}}={1 \over \Gamma(\nu)} {1 \over 2\pi i} \int_{-i\infty}^{+i\infty}ds\,
\Gamma(-s)\Gamma(\nu+s){B^s \over A^{\nu+s}}\:.
\label{MB-1}
\end{eqnarray}
Then, one can have
\begin{eqnarray}
&&{1 \over (k_{_i}^2-m_{_i}^2)^{\nu_{_i}}}={1 \over \Gamma(\nu_{_i})} {1 \over 2\pi i} \int_{-i\infty}^{+i\infty}ds_{_i}\,
\Gamma(-s_{_i})\Gamma(\nu_{_i}+s_{_i}){(-m_{_i}^2)^{s_{_i}} \over (k_{_i}^2)^{\nu_{_i}+s_{_i}}}\:.
\label{MB-2}
\end{eqnarray}

It's relatively easy to compute analytically momentum integral when most virtual masses are zero. And then, one can give the Mellin-Barnes representation of Feynman integral
\begin{eqnarray}
U_{_N}\propto \int_{-i\infty}^{+i\infty}d{\bf s}\, f({\bf s}) \Big[\prod\limits_{i=1}^M x_{_i}^{s_{_i}}\Gamma(-s_{_i}) \Big]\:,
\label{MB-3}
\end{eqnarray}
where ${\bf s}=(s_{_1},\,\cdots,\,s_{_M})$ with that number of variables  ${\bf s}$ is $M$, $d{\bf s}=\prod\limits_{i=1}^M ds_{_i}$, $f({\bf s})$ denotes a function including gamma functions with variables $\bf s$, $x_{_i}$ denotes the ratio of virtual mass squared and external momentum squared.

It is well known that zero and negative integers are simple poles of the gamma function
$\Gamma(z)$. As all $s_{_i}$ contours are closed to the right in
complex planes, one can find that the analytic expression of Feynman  integral
can be written as the linear combination of generalized hypergeometric functions. Taking the residue of the pole of $\Gamma(-s_{_i}),\;(i=1,\cdots,M)$, we can derive one linear independent term of Feynman integral:
\begin{eqnarray}
U_{_N}\ni T_{_{N}}({\bf a},\;{\bf b}\;\Big|\;{\bf x})\:,
\label{MB-4}
\end{eqnarray}
where $T_{_{N}}$ is a hypergeometric function with parameters ${\bf a}$ and ${\bf b}$. Here, we just derive one linear independent term of the integral. We still need to derive other linear independent terms of the integral. Next, we can construct GKZ hypergeometric system of Feynman integral to obtain the other linear independent terms of the integral.

We define the auxiliary function
\begin{eqnarray}
&&\Phi_{_{N}}({\bf a},\;{\bf b}\;\Big|\;{\bf x},\;{\bf u},\;{\bf v})={\bf u}^{\bf a}{\bf v}^{{\bf b}-{\bf e}_{_K}}T_{_{N}}({\bf a},\;{\bf b}\;\Big|\;{\bf x})\;,
\label{GKZ-1}
\end{eqnarray}
with the intermediate variables ${\bf u}=(u_{_1},\,\cdots,\,u_{_J})$ and ${\bf v}=(v_{_1},\,\cdots,\,v_{_K})$, $J$ is number of parameters ${\bf a}$ and $K$ is number of parameters ${\bf b}$, and ${\bf e}_{_K}=(1,\,\cdots,\,1)$ with that number of 1 is $K$.
Through Miller's transformation \cite{Miller68,Miller72}, one can have the relations
\begin{eqnarray}
&&\vartheta_{_{u_j}}\Phi_{_{N}}({\bf a},\;{\bf b}\;\Big|\;{\bf x},\;{\bf u},\;{\bf v})=a_{_j}
\Phi_{_{N}}({\bf a},\;{\bf b}\;\Big|\;{\bf x},\;{\bf u},\;{\bf v})
\;,(j=1,\cdots,J)\:,\nonumber\\
&&\vartheta_{_{v_k}}\Phi_{_{N}}({\bf a},\;{\bf b}\;\Big|\;{\bf x},\;{\bf u},\;{\bf v})=(b_{_k}-1)
\Phi_{_{N}}({\bf a},\;{\bf b}\;\Big|\;{\bf x},\;{\bf u},\;{\bf v})\;,(k=1,\cdots,K)\:,
\label{GKZ-2}
\end{eqnarray}
where $\vartheta_{_{x_{_i}}}=x_{_i}\partial_{_{x_{_i}}}$
denotes the Euler operators, and $\partial_{_{x_{_i}}}=\partial/\partial x_{_i}$. The relations naturally induce the notion of GKZ hypergeometric system.

Through the appropriate transformation of variables ${\bf u},\,{\bf v},{\bf x}$ to variables ${\bf z}$,
we can finally have the GKZ hypergeometric system of Feynman integral:
\begin{eqnarray}
&&\mathbf{A_{_{N}}}\cdot\vec{\vartheta}_{_{N}}\Phi_{_{N}}=\mathbf{B_{_{N}}}\Phi_{_{N}}\;,
\label{GKZ-3}
\end{eqnarray}
where the expressions of $\mathbf{A_{_{N}}}$, $\vec{\vartheta}_{_{N}}$, $\mathbf{B_{_{N}}}$ depend on the given Feynman integral. In the following section, we take 4-loop vacuum integral with five propagates as an example to expand concretely.

\section{GKZ hypergeometric system of 4-loop vacuum integral with five propagates\label{sec-five}}

\begin{figure}[th]
\setlength{\unitlength}{0cm}
\centering
\hspace{-1.cm}\hspace{2cm}
\includegraphics[width=7cm]{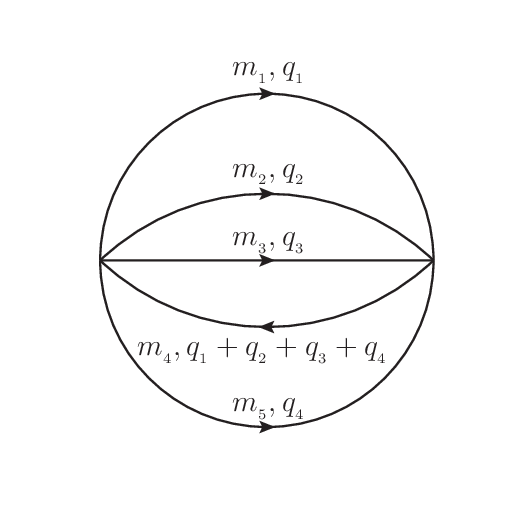}
\vspace{-1.0cm}
\caption[]{Four-loop vacuum diagram with five propagators, which $m_{_i}$ denotes the mass of the $i$-th particle and $q_{_j}$ denotes the loop momentum.}
\label{fig-4loop-1q}
\end{figure}

In Fig.~\ref{fig-4loop-1q}, we picture 4-loop vacuum diagram with five propagates, which is simplest 4-loop vacuum diagram. The diagram looks like a banana, so we name it as banana-type.
The expression of banana-type 4-loop vacuum integral can be written as
\begin{eqnarray}
&&U_{_5}=\Big(\Lambda_{_{\rm RE}}^2\Big)^{8-2D}\int{d^D{\bf q}\over(2\pi)^D}
{1\over(q_{_1}^2-m_{_1}^2)^{\nu_{_1}}(q_{_2}^2-m_{_2}^2)^{\nu_{_2}}(q_{_3}^2-m_{_3}^2)^{\nu_{_3}}}
\nonumber\\
&&\hspace{1.1cm}\times
{1\over [(q_{_1}+q_{_2}+q_{_3}+q_{_4})^2-m_{_4}^2]^{\nu_{_4}}
(q_{_4}^2-m_{_5}^2)^{\nu_{_5}}}\:,
\label{4loop-1}
\end{eqnarray}
Through Mellin-Barnes transformation, the four-loop vacuum integral is written as
\begin{eqnarray}
&&U_{_5}=\frac{\Big(\Lambda_{_{\rm RE}}^2\Big)^{8-2D}}{(2\pi i)^4 \prod\limits_{i=1}^4\Gamma(\nu_{_i})} \int_{-i\infty}^{+i\infty}d{\bf s}
\Big[\prod\limits_{i=1}^4(-m_{_i}^2)^{s_{_i}}\Gamma(-s_{_i})\Gamma(\nu_{_i}+s_{_i})\Big]I_{q}\:,
\label{4loop-2}
\end{eqnarray}
where ${\bf s}=(s_{_1},\;\cdots,\;s_{_4})$, $d{\bf s}=ds_{_1}ds_{_2}ds_{_3}ds_{_4}$ and
\begin{eqnarray}
&& I_{q}\equiv \int{d^D{\bf q}\over(2\pi)^D}
{1\over(q_{_1}^2)^{\nu_{_1}+s_{_1}}(q_{_2}^2)^{\nu_{_2}+s_{_2}}(q_{_3}^2)^{\nu_{_3}+s_{_3}}
[(q_{_1}+q_{_2}+q_{_3}+q_{_4})^2]^{\nu_{_4}+s_{_4}}
(q_{_4}^2-m_{_5}^2)^{\nu_{_5}}}\:.
\label{4loop-2-1}
\end{eqnarray}
The integral $I_{q}$ just keeps one mass $m_{_5}$, which can be easily calculated analytically.
One can have
\begin{eqnarray}
I_{q}&&={{1} \over (4\pi)^{2D}\Gamma(\nu_{_5})} (-)^{\sum\limits_{i=1}^5 \nu_{_i}+\sum\limits_{i=1}^4s_{_i}} \Big( {1\over m_{_5}^2}\Big)^{\sum\limits_{i=1}^5 \nu_{_i}-2D+\sum\limits_{i=1}^4s_{_i}}
\Big[\prod\limits_{i=1}^4\Gamma({D\over2}-\nu_{_i}-s_{_i})\Gamma(\nu_{_i}+s_{_i})^{-1}\Big]
\nonumber\\
&&\hspace{0.5cm}\times
\Gamma(\sum\limits_{i=1}^4 \nu_{_i}-{3D\over2}+\sum\limits_{i=1}^4s_{_i})\Gamma(\sum\limits_{i=1}^5 \nu_{_i}-2D+\sum\limits_{i=1}^4s_{_i})
\;.
\label{Iq-7-1}
\end{eqnarray}
The detailed calculation of the integral $I_{q}$ can be seen in Appendix~\ref{app-Iq}.

And then, the Mellin-Barnes representation of the four-loop vacuum integral in Eq. (\ref{4loop-2}) can be written as
\begin{eqnarray}
&&U_{_5}=
{1\over(2\pi i)^4(4\pi)^8 \prod\limits_{i=1}^5\Gamma(\nu_{_i})}(-)^{\sum\limits_{i=1}^5 \nu_{_i}}
(m_{_5}^2)^{8-\sum\limits_{i=1}^5 \nu_{_i}} \Big({4\pi\Lambda_{_{\rm RE}}^2\over m_{_5}^2}\Big)^{8-2D}
\int_{-i\infty}^{+i\infty}d{\bf s}
\Big[\prod\limits_{i=1}^4\Big({m_{_i}^2\over m_{_5}^2}\Big)^{s_{_i}}\Gamma(-s_{_i})\Big]
\nonumber\\
&&\hspace{1.0cm}\times
\Big[\prod\limits_{i=1}^4\Gamma({D\over2}-\nu_{_i}-s_{_i})\Big]
\Gamma(\sum\limits_{i=1}^4 \nu_{_i}-{3D\over2}+\sum\limits_{i=1}^4s_{_i})
\Gamma(\sum\limits_{i=1}^5 \nu_{_i}-2D+\sum\limits_{i=1}^4s_{_i})
\;.
\label{4loop-7}
\end{eqnarray}

Taking the residue of the pole of $\Gamma(-s_{_i}),\;(i=1,\cdots,4)$, we derive one linear independent term of the integral:
\begin{eqnarray}
&&\hspace{-0.5cm}U_{_5}\ni
{1\over (4\pi)^8\prod\limits_{i=1}^5\Gamma(\nu_{_i})} (-)^{\sum\limits_{i=1}^5 \nu_{_i}}
(m_{_5}^2)^{8-\sum\limits_{i=1}^5 \nu_{_i}} \Big({4\pi\Lambda_{_{\rm RE}}^2\over m_{_5}^2}\Big)^{8-2D}
\sum\limits_{n_{_1}=0}^\infty\sum_{n_{_2}=0}^\infty\sum_{n_{_3}=0}^\infty \sum_{n_{_4}=0}^\infty
(-)^{\sum\limits_{i=1}^4 n_{_i}}x_{_1}^{n_{_1}}x_{_2}^{n_{_2}}x_{_3}^{n_{_3}}x_{_4}^{n_{_4}}
\nonumber\\
&&\hspace{0.5cm}\times
\Big[\prod\limits_{i=1}^4\Gamma({D\over2}-\nu_{_i}-n_{_i})(n_{_i}!)^{-1}\Big]
\Gamma(\sum\limits_{i=1}^4 \nu_{_i}-{3D\over2}+\sum\limits_{i=1}^4n_{_i})\Gamma(\sum\limits_{i=1}^5 \nu_{_i}-2D+\sum\limits_{i=1}^4n_{_i})
\:,
\label{4loop-8}
\end{eqnarray}
with $x_{_i}={m_{_i}^2\over m_{_5}^2}$.

We adopt the well-known identity
\begin{eqnarray}
&&\Gamma(z-n)\Gamma(1-z+n)=(-)^{n}\Gamma(z)\Gamma(1-z)=(-)^{n}\pi/\sin\pi z\:.
\label{Gamma}
\end{eqnarray}
Then, Eq.~(\ref{4loop-8}) is written as
\begin{eqnarray}
U_{_5}\ni
{1\over (4\pi)^8\prod\limits_{i=1}^5\Gamma(\nu_{_i})} (-)^{ \nu_{_5}}
(m_{_5}^2)^{8-\sum\limits_{i=1}^5 \nu_{_i}} \Big({4\pi\Lambda_{_{\rm RE}}^2\over m_{_5}^2}\Big)^{8-2D}
{ \pi^4\over \sin^4\frac{\pi D}{2} }
T_{_{5}}({\bf a},\;{\bf b}\;\Big|\;{\bf x})\;,
\label{4loop-9}
\end{eqnarray}
with a hypergeometric function
\begin{eqnarray}
T_{_{5}}({\bf a},\;{\bf b}\;\Big|\;{\bf x})=\sum\limits_{n_{_1}=0}^\infty\sum_{n_{_2}=0}^\infty\sum_{n_{_3}=0}^\infty\sum_{n_{_4}=0}^\infty
A_{_{n_{_1}n_{_2}n_{_3}n_{_4}}} x_{_1}^{n_{_1}}x_{_2}^{n_{_2}}x_{_3}^{n_{_3}}x_{_4}^{n_{_4}}\:,
\label{4loop-10}
\end{eqnarray}
where ${\bf x}=(x_{_1},x_{_2},x_{_3},x_{_4})$, ${\bf a}=(a_{_1},a_{_2})$ and ${\bf b}=(b_{_1},b_{_2},b_{_3},b_{_4})$ with
\begin{eqnarray}
&&a_{_1}=\sum\limits_{i=1}^4 \nu_{_i}-{3D\over2},\;\;a_{_2}=\sum\limits_{i=1}^5 \nu_{_i}-2D,\;\;
b_{_i}=1+\nu_{_i}-{D\over2}\:,
\label{4loop-11}
\end{eqnarray}
and the coefficient $A_{_{n_{_1}n_{_2}n_{_3}n_{_4}}}$ is
\begin{eqnarray}
\hspace{-0.5cm}A_{_{n_{_1}n_{_2}n_{_3}n_{_4}}}=
{\Gamma(a_{_1}+\sum\limits_{i=1}^4n_{_i}) \Gamma(a_{_2}+\sum\limits_{i=1}^4n_{_i})
\over
\prod\limits_{i=1}^4 n_{_i}! \Gamma(b_{_i}+n_{_i})}\:.
\label{4loop-12}
\end{eqnarray}

Through the reduction of 4-loop vacuum integrals in reference \cite{R4loop4}, the numerators of the master integrals for 4-loop vacuum diagrams are equal to 1, which can be seen figure 3 of the reference. And the denominators powers of the master integrals for 4-loop vacuum diagram with five propagates have two cases: $\nu_{_1}=1$ and $\nu_{_1}=3$ with that other powers are equal to 1. Here, for  $\nu_{_1}=1$ case, the parameters ${\bf a}$ and ${\bf b}$ are
\begin{eqnarray}
&&a_{_1}=4-{3D\over2},\;\;a_{_2}=5-2D,\;\;
b_{_i}=2-{D\over2},\;(i=1,\cdots,4)\:,
\label{4loop-11-1}
\end{eqnarray}
For  $\nu_{_1}=3$ case, the parameters ${\bf a}$ and ${\bf b}$ are
\begin{eqnarray}
&&a_{_1}=6-{3D\over2},\;\;a_{_2}=7-2D,\;\;
b_{_1}=4-{D\over2},\;\;b_{_i}=2-{D\over2},\;(i=2,3,4)\:.
\label{4loop-11-2}
\end{eqnarray}

We define the auxiliary function
\begin{eqnarray}
&&\Phi_{_{5}}({\bf a},\;{\bf b}\;\Big|\;{\bf x},\;{\bf u},\;{\bf v})={\bf u}^{\bf a}{\bf v}^{{\bf b}-{\bf e}_{_{4}}}T_{_{5}}({\bf a},\;{\bf b}\;\Big|\;{\bf x})\;,
\label{4loop-15}
\end{eqnarray}
with the intermediate variables ${\bf u}=(u_{_1},\;u_{_2})=(1,\;1)$, ${\bf v}=(v_{_1},\;v_{_2},\;v_{_3},\;v_{_4})=(1,\;1,\;1,\;1)$, ${\bf e}_{_{4}}=(1,\;1,\;1,\;1)$.
Through Miller's transformation, one can be obtained
\begin{eqnarray}
&&\vartheta_{_{u_j}}\Phi_{_{5}}({\bf a},\;{\bf b}\;\Big|\;{\bf x},\;{\bf u},\;{\bf v})=a_{_j}
\Phi_{_{5}}({\bf a},\;{\bf b}\;\Big|\;{\bf x},\;{\bf u},\;{\bf v})
\;,\;(j=1,2)\:,\nonumber\\
&&\vartheta_{_{v_k}}\Phi_{_{5}}({\bf a},\;{\bf b}\;\Big|\;{\bf x},\;{\bf u},\;{\bf v})=(b_{_k}-1)
\Phi_{_{5}}({\bf a},\;{\bf b}\;\Big|\;{\bf x},\;{\bf u},\;{\bf v})\;,\;(k=1,\cdots,4)\:,
\label{4loop-16}
\end{eqnarray}
which naturally induces the notion of GKZ hypergeometric system.

Through the transformation
\begin{eqnarray}
&&z_{_j}={1\over u_{_j}},\;\;z_{_{2+k}}=v_{_k},\;z_{_{6+k}}={x_{_k}\over u_{_1}u_{_2}v_{_k}},
\label{4loop-18}
\end{eqnarray}
one can finally have the GKZ hypergeometric system for the four-loop vacuum integral  with five propagates:
\begin{eqnarray}
&&\mathbf{A_{_{5}}}\cdot\vec{\vartheta}_{_{5}}\Phi_{_{5}}=\mathbf{B_{_{5}}}\Phi_{_{5}}\;,
\label{4loop-20}
\end{eqnarray}
where
\begin{eqnarray}
&&\mathbf{A_{_{5}}}=\left(\begin{array}{cccccccccc}
1\;\;&0\;\;&0\;\;&0\;\;&0\;\;&0\;\;&1\;\;&1\;\;&1\;\;&1\;\;\\
0\;\;&1\;\;&0\;\;&0\;\;&0\;\;&0\;\;&1\;\;&1\;\;&1\;\;&1\;\;\\
0\;\;&0\;\;&1\;\;&0\;\;&0\;\;&0\;\;&-1\;\;&0\;\;&0\;\;&0\;\;\\
0\;\;&0\;\;&0\;\;&1\;\;&0\;\;&0\;\;&0\;\;&-1\;\;&0\;\;&0\;\;\\
0\;\;&0\;\;&0\;\;&0\;\;&1\;\;&0\;\;&0\;\;&0\;\;&-1\;\;&0\;\;\\
0\;\;&0\;\;&0\;\;&0\;\;&0\;\;&1\;\;&0\;\;&0\;\;&0\;\;&-1\;\;\\
\end{array}\right)
\;,\nonumber\\
&&\vec{\vartheta}_{_{5}}^{\;T}=(\vartheta_{_{z_{_1}}},\;\cdots
,\;\vartheta_{_{z_{_{10}}}})
\;,\nonumber\\
&&\mathbf{B_{_{5}}}^{T}=(-a_{_1},\;-a_{_2},\;b_{_1}-1,\;b_{_2}-1
,\;b_{_3}-1,\;b_{_4}-1)\;.
\label{4loop-21}
\end{eqnarray}

Correspondingly the dual matrix $\mathbf{\tilde A_{_{5}}}$ of $\mathbf{A_{_{5}}}$ is
\begin{eqnarray}
&&\mathbf{\tilde A_{_{5}}}=\left(\begin{array}{cccccccccc}
-1\;\;&-1\;\;&1\;\;&0\;\;&0\;\;&0\;\;&1\;\;&0\;\;&0\;\;&0\;\;\\
-1\;\;&-1\;\;&0\;\;&1\;\;&0\;\;&0\;\;&0\;\;&1\;\;&0\;\;&0\;\;\\
-1\;\;&-1\;\;&0\;\;&0\;\;&1\;\;&0\;\;&0\;\;&0\;\;&1\;\;&0\;\;\\
-1\;\;&-1\;\;&0\;\;&0\;\;&0\;\;&1\;\;&0\;\;&0\;\;&0\;\;&1\;\;
\end{array}\right).
\label{4loop-22}
\end{eqnarray}
The row vectors of the matrix $\mathbf{\tilde A_{_{5}}}$ induce the integer sublattice $\mathbf{B}$
which can be used to construct the formal solutions in hypergeometric series.

Defining the combined variables
\begin{eqnarray}
y_{_1}={z_{_3}z_{_{7}}\over z_{_1}z_{_2}}
\;,\quad  y_{_2}={z_{_4}z_{_{8}}\over z_{_1}z_{_2}}
\;,\quad  y_{_3}={z_{_5}z_{_{9}}\over z_{_1}z_{_2}}
\;,\quad  y_{_4}={z_{_6}z_{_{10}}\over z_{_1}z_{_2}}
\;,
\label{4loop-24}
\end{eqnarray}
we can write the solutions satisfying Eq. (\ref{4loop-20}) as
\begin{eqnarray}
&&\Phi_{_{5}}({\mathbf z})=\Big(\prod\limits_{i=1}^{10}z_{_i}^{\alpha_{_i}}\Big)
\varphi_{_{5}}(y_{_1},\;y_{_2},\;y_{_3},\;y_{_4})\;.
\label{4loop-25}
\end{eqnarray}
Here $\vec{\alpha}^{\:T}=(\alpha_{_1},\;\cdots,\;\alpha_{_{10}})$
denotes a sequence of complex number such that
\begin{eqnarray}
&&\mathbf{A_{_{5}}}\cdot\vec{\alpha}=\mathbf{B_{_{5}}}\;.
\label{4loop-26}
\end{eqnarray}
In the following Sec. \ref{sec-solu}, we will show the analytical hypergeometric series solutions solving from the GKZ hypergeometric system in Eq. (\ref{4loop-20}).

\section{GKZ hypergeometric system of 4-loop vacuum integrals with six propagates\label{sec-six}}

\subsection{4-loop vacuum diagram with six propagates for flower-type}

\begin{figure}[t]
\setlength{\unitlength}{0cm}
\centering
\vspace{-1cm}\hspace{1cm}
\includegraphics[width=7.8cm]{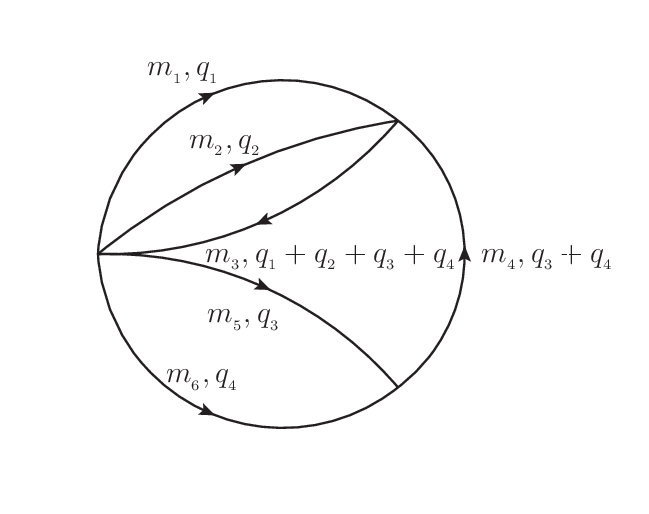}
\vspace{-1cm}
\caption[]{Four-loop vacuum diagram with six propagators for flower-type.}
\label{fig-4loop-2q}
\end{figure}

The four-loop vacuum diagrams with six propagates have two topologies, which can be seen in Fig.~\ref{fig-4loop-2q} and Fig.~\ref{fig-4loop-3q}. In Fig.~\ref{fig-4loop-2q}, the four-loop vacuum diagram just likes a beautiful flower, so we name it as flower-type.
The four-loop vacuum diagram with six propagates for flower-type just has one master integral, which is
\begin{eqnarray}
&&U_{_{6A}}=\Big(\Lambda_{_{\rm RE}}^2\Big)^{8-2D}\int{d^D{\bf q}\over(2\pi)^D}
{1\over(q_{_1}^2-m_{_1}^2)(q_{_2}^2-m_{_2}^2)[(q_{_1}+q_{_2}+q_{_3}+q_{_4})^2-m_{_3}^2]}
\nonumber\\
&&\hspace{1.1cm}\times
{1\over
[(q_{_3}+q_{_4})^2-m_{_4}^2](q_{_3}^2-m_{_5}^2)(q_{_4}^2-m_{_6}^2)}.
\label{GKZ6A0}
\end{eqnarray}

Similarly, the Mellin-Barnes representation of the flower-type four-loop vacuum integral can be written as
\begin{eqnarray}
&&U_{_{6A}}=
{m_{_6}^4\over(2\pi i)^5(4\pi)^8}\Big({4\pi\Lambda_{_{\rm RE}}^2\over m_{_6}^2}\Big)^{8-2D}
\int_{-i\infty}^{+i\infty}d{\bf s}\Big[\prod\limits_{i=1}^5\Big({m_{_i}^2\over m_{_6}^2}\Big)^{s_{_i}}\Gamma(-s_{_i})\Big]
\nonumber\\
&&\hspace{1.0cm}\times
\Big[\prod\limits_{i=1}^3\Gamma({D\over2}-1-s_{_i})\Big]\Gamma({D\over2}-1-s_{_5})\Gamma(5-{3D\over2}+\sum\limits_{i=1}^5s_{_i})
\nonumber\\
&&\hspace{1.0cm}\times
{\Gamma(6-2D+\sum\limits_{i=1}^5s_{_i})\Gamma(3-D+\sum\limits_{i=1}^3s_{_i})\Gamma({3D\over2}-4-\sum\limits_{i=1}^4s_{_i})\Gamma(1+s_{_4})
\over\Gamma(4-D+\sum\limits_{i=1}^4s_{_i})\Gamma({3D\over2}-3-\sum\limits_{i=1}^3s_{_i})}\;,
\label{GKZ6A2}
\end{eqnarray}
where  ${\bf s}=(s_{_1},\;\cdots,\;s_{_5})$.

Taking the residue of the pole of $\Gamma(-s_{_i}),\;(i=1,\cdots,5)$, we can derive one linear independent term of the integral:
\begin{eqnarray}
&&U_{_{6A}}\ni
{m_{_6}^4\over(4\pi)^8}\Big({4\pi\Lambda_{_{\rm RE}}^2\over m_{_6}^2}\Big)^{8-2D}
{ \pi^4\over \sin^4\frac{\pi D}{2} }
T_{_{6A}}({\bf a},\;{\bf b}\;\Big|\;{\bf x})\;,
\label{GKZ6A5}
\end{eqnarray}
with
\begin{eqnarray}
T_{_{6A}}({\bf a},\;{\bf b}\;\Big|\;{\bf x})={\bf{\sum\limits_{{n}=0}^\infty}} A_{_{\bf n}} {\bf x^{n}},
\label{GKZ6A6}
\end{eqnarray}
where ${\bf n}=(n_{_1},\;\cdots,\;n_{_5})$, ${\bf x}=(x_{_1},\;\cdots,\;x_{_5})$, $x_{_i}={m_{_i}^2\over m_{_6}^2}$, ${\bf a}=(a_{_1},\;\cdots,a_{_5})$ and
${\bf b}=(b_{_1},\;\cdots,b_{_6})$ with
\begin{eqnarray}
&&a_{_1}=5-{3D\over2},\;a_{_2}=6-2D,\;a_{_3}=3-D,\;
a_{_4}=4-{3D\over2},\;a_{_5}=1,
\nonumber\\
&&b_{_1}=b_{_2}=b_{_3}=b_{_4}=2-{D\over2},\;b_{_5}=4-D,\;b_{_6}=5-{3D\over2}\:,
\label{GKZ6A6-1}
\end{eqnarray}
and the coefficient $A_{_{\bf n}}$ is
\begin{eqnarray}
\hspace{-0.5cm}A_{_{{\bf n}}}=
{\Gamma(a_{_1}+\sum\limits_{i=1}^5n_{_i}) \Gamma(a_{_2}+\sum\limits_{i=1}^5n_{_i})\Gamma(a_{_3}+\sum\limits_{i=1}^3n_{_i})\Gamma(a_{_4}+\sum\limits_{i=1}^3n_{_i})\Gamma(a_{_5}+n_{_4})
\over  \Big[\prod\limits_{i=1}^5 n_{_i}!\Big] \Big[\prod\limits_{i=1}^3\Gamma(b_{_i}+n_{_i})\Big]\Gamma(b_{_4}+n_{_5})
\Gamma(b_{_5}+\sum\limits_{i=1}^4n_{_i})\Gamma(b_{_6}+\sum\limits_{i=1}^4n_{_i})}.
\label{GKZ6A6-2}
\end{eqnarray}

We also define the auxiliary function
\begin{eqnarray}
&&\Phi_{_{6A}}({\bf a},\;{\bf b}\;\Big|\;{\bf x},\;{\bf u},\;{\bf v})={\bf u}^{\bf a}{\bf v}^{{\bf b}-{\bf e}_{_{6}}}
T_{_{6A}}({\bf a},\;{\bf b}\;\Big|\;{\bf x})\;,
\label{GKZ6A10}
\end{eqnarray}
with the intermediate variables ${\bf u}=(u_{_1},\cdots,u_{_5})=(1,\;1,\;1,\;1,\;1)$, ${\bf v}=(v_{_1},\cdots,v_{_6})$, ${\bf v}={\bf e}_{_{6}}=(1,\;1,\;1,\;1,\;1,\;1)$.
One can derive the GKZ hypergeometric system for the four-loop vacuum integral for flower-type:
\begin{eqnarray}
&&\mathbf{A_{_{6A}}}\cdot\vec{\vartheta}_{_{6A}}\Phi_{_{6A}}=\mathbf{B_{_{6A}}}\Phi_{_{6A}}\;,
\label{GKZ6A17}
\end{eqnarray}
where
\begin{eqnarray}
&&\mathbf{A_{_{6A}}}=\left(\begin{array}{cc}
\mathbf{I_{_{11\times11}}}\;\;&\mathbf{A_{_{X6A}}}\;\;\\
\end{array}\right)
\;,\nonumber\\
&&\mathbf{A_{_{X6A}}}^{\;T}=\left(\begin{array}{ccccccccccc}
1\;\;&1\;\;&1\;\;&1\;\;&0\;\;&-1\;\;&0\;\;&0\;\;&0\;\;&-1\;\;&-1\;\;\\
1\;\;&1\;\;&1\;\;&1\;\;&0\;\;&0\;\;&-1\;\;&0\;\;&0\;\;&-1\;\;&-1\;\;\\
1\;\;&1\;\;&1\;\;&1\;\;&0\;\;&0\;\;&0\;\;&-1\;\;&0\;\;&-1\;\;&-1\;\;\\
1\;\;&1\;\;&0\;\;&0\;\;&1\;\;&0\;\;&0\;\;&0\;\;&0\;\;&-1\;\;&-1\;\;\\
1\;\;&1\;\;&0\;\;&0\;\;&0\;\;&0\;\;&0\;\;&0\;\;&-1\;\;&0\;\;&0\;\;\\
\end{array}\right)
\;,\nonumber\\
&&\vec{\vartheta}_{_{6A}}^{\;T}=(\vartheta_{_{z_{_1}}},\cdots,\;\vartheta_{_{z_{_{16}}}})
\;,\nonumber\\
&&\mathbf{B_{_{6A}}}^{\;T}=(-a_{_1},\;\cdots,\;-a_{_5},\;b_{_1}-1,\;\cdots,\;b_{_6}-1)\;.
\label{GKZ6A18}
\end{eqnarray}
Here, $\mathbf{I_{_{11\times11}}}$ is an ${11\times11}$ unit matrix.
Correspondingly the dual matrix $\mathbf{\tilde A_{_{6A}}}$ of $\mathbf{A_{_{6A}}}$ is
\begin{eqnarray}
&&\mathbf{\tilde A_{_{6A}}}=\left(\begin{array}{cc}
-\mathbf{A_{_{X6A}}}^{\;T}\;\;&\mathbf{I_{_{5\times5}}}\;\;\\
\end{array}\right).
\label{GKZ6A19}
\end{eqnarray}
Here, $\mathbf{I_{_{5\times5}}}$ is a ${5\times5}$ unit matrix.

\subsection{4-loop vacuum diagram with six propagates for triangle-type}

\begin{figure}[t]
\setlength{\unitlength}{0cm}
\centering
\vspace{-1cm}\hspace{1cm}
\includegraphics[width=9.0cm]{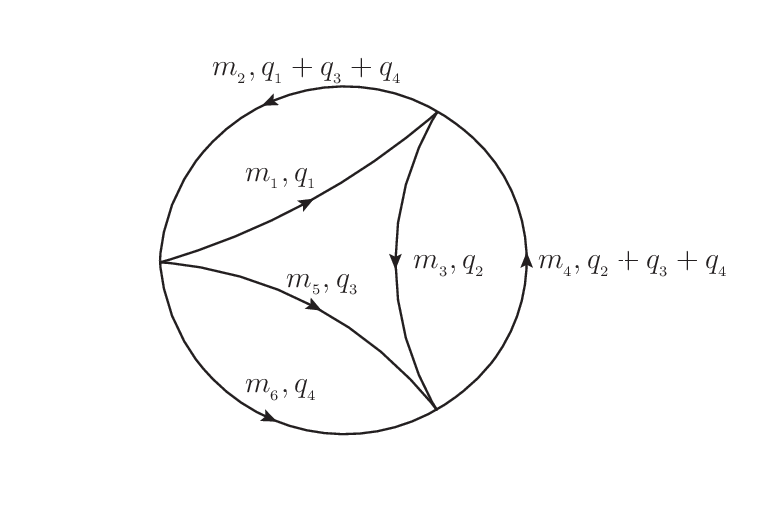}
\vspace{-1cm}
\caption[]{Four-loop vacuum diagram with six propagators for triangle-type.}
\label{fig-4loop-3q}
\end{figure}

In Fig.~\ref{fig-4loop-3q}, we show the other four-loop vacuum diagram with six propagates. The diagram looks like a triangle inside except for the outer circle, so we name it as triangle-type.
The master integral of the four-loop vacuum diagram with six propagates for triangle-type can be written as
\begin{eqnarray}
&&U_{_{6B}}=\Big(\Lambda_{_{\rm RE}}^2\Big)^{8-2D}\int{d^D{\bf q}\over(2\pi)^D}
{1 \over (q_{_1}^2-m_{_1}^2)[(q_{_1}+q_{_3}+q_{_4})^2-m_{_2}^2]}
\nonumber\\
&&\hspace{1.1cm}\times
{1\over (q_{_2}^2-m_{_3}^2)
[(q_{_2}+q_{_3}+q_{_4})^2-m_{_4}^2](q_{_3}^2-m_{_5}^2)(q_{_4}^2-m_{_6}^2)}.
\label{GKZ6B0}
\end{eqnarray}
Integrating out ${\bf q}$, the Mellin-Barnes representation of the four-loop vacuum integral for triangle-type is written as
\begin{eqnarray}
&&U_{_{6B}}=
{m_{_6}^4\over(2\pi i)^5(4\pi)^8}\Big({4\pi\Lambda_{_{\rm RE}}^2\over m_{_6}^2}\Big)^{8-2D}
\int_{-i\infty}^{+i\infty}d{\bf s}\Big[\prod\limits_{i=1}^5\Big({m_{_i}^2\over m_{_6}^2}\Big)^{s_{_i}}\Gamma(-s_{_i})\Big]
\nonumber\\
&&\hspace{1.0cm}\times
\Big[\prod\limits_{i=1}^5\Gamma({D\over2}-1-s_{_i})\Big]\Gamma(5-{3D\over2}+\sum\limits_{i=1}^5s_{_i})
\Gamma(6-2D+\sum\limits_{i=1}^5s_{_i})\nonumber\\
&&\hspace{1.0cm}\times
{\Gamma({3D\over2}-4-\sum\limits_{i=1}^4s_{_i})
\Gamma(2-{D\over2}+s_{_1}+s_{_2})\Gamma(2-{D\over2}+s_{_3}+s_{_4})
\over\Gamma(4-D+\sum\limits_{i=1}^4s_{_i})\Gamma(D-2-s_{_1}-s_{_2})\Gamma(D-2-s_{_3}-s_{_4})}\;.
\label{GKZ6B2}
\end{eqnarray}

And then, we can derive one linear independent term of the integral:
\begin{eqnarray}
&&U_{_{6B}}\ni
{m_{_6}^4\over(4\pi)^8}\Big({4\pi\Lambda_{_{\rm RE}}^2\over m_{_6}^2}\Big)^{8-2D}
{ \pi^4 \sin^2\pi D \over \sin^5\frac{\pi D}{2} \sin\frac{3\pi D}{2} }
T_{_{6B}}({\bf a},\;{\bf b}\;\Big|\;{\bf x})\;,
\label{GKZ6B5}
\end{eqnarray}
with
\begin{eqnarray}
T_{_{6B}}({\bf a},\;{\bf b}\;\Big|\;{\bf x})={\bf{\sum\limits_{{n}=0}^\infty}} A_{_{\bf n}} {\bf x^{n}},
\label{GKZ6B6}
\end{eqnarray}
where ${\bf a}=(a_{_1},\;\cdots,a_{_6})$ and
${\bf b}=(b_{_1},\;\cdots,b_{_7})$ with
\begin{eqnarray}
&&a_{_1}=5-{3D\over2},\;a_{_2}=6-2D,\;a_{_3}=a_{_5}=2-{D\over2},\;
a_{_4}=a_{_6}=3-D\;,
\nonumber\\
&&b_{_1}=b_{_2}=b_{_3}=b_{_4}=b_{_5}=2-{D\over2},\;b_{_6}=4-D,\;b_{_7}=5-{3D\over2}\:,
\label{GKZ6B6-1}
\end{eqnarray}
and the coefficient $A_{_{\bf n}}$ is
\begin{eqnarray}
\hspace{-0.5cm}A_{_{{\bf n}}}=
{\Gamma(a_{_1}+\sum\limits_{i=1}^5n_{_i}) \Gamma(a_{_2}+\sum\limits_{i=1}^5n_{_i})
\Gamma(a_{_3}+\sum\limits_{i=1}^2n_{_i})\Gamma(a_{_4}+\sum\limits_{i=1}^2n_{_i})
\Gamma(a_{_5}+\sum\limits_{i=3}^4n_{_i})\Gamma(a_{_6}+\sum\limits_{i=3}^4n_{_i})
\over  \Big[\prod\limits_{i=1}^5 n_{_i}! \Gamma(b_{_i}+n_{_i})\Big]
\Gamma(b_{_6}+\sum\limits_{i=1}^4n_{_i})\Gamma(b_{_7}+\sum\limits_{i=1}^4n_{_i})}.
\label{GKZ6B6-2}
\end{eqnarray}

We can define the auxiliary function
\begin{eqnarray}
&&\Phi_{_{6B}}({\bf a},\;{\bf b}\;\Big|\;{\bf x},\;{\bf u},\;{\bf v})={\bf u}^{\bf a}{\bf v}^{{\bf b}-{\bf e}_{_{7}}}
T_{_{6B}}({\bf a},\;{\bf b}\;\Big|\;{\bf x})\;,
\label{GKZ6B10}
\end{eqnarray}
with the intermediate variables ${\bf u}=(u_{_1},\cdots,u_{_6})=(1,\;1,\;1,\;1,\;1,\;1)$, ${\bf v}=(v_{_1},\cdots,v_{_7})$, ${\bf v}={\bf e}_{_{7}}=(1,\;1,\;1,\;1,\;1,\;1,\;1)$.
One can give the GKZ hypergeometric system for the four-loop vacuum integral for triangle-type:
\begin{eqnarray}
&&\mathbf{A_{_{6B}}}\cdot\vec{\vartheta}_{_{6B}}\Phi_{_{6B}}=\mathbf{B_{_{6B}}}\Phi_{_{6B}}\;,
\label{GKZ6B17}
\end{eqnarray}
where
\begin{eqnarray}
&&\mathbf{A_{_{6B}}}=\left(\begin{array}{cc}
\mathbf{I_{_{13\times13}}}\;\;&\mathbf{A_{_{X6B}}}\;\;\\
\end{array}\right)
\;,\nonumber\\
&&\mathbf{A_{_{X6B}}}^{\;T}=\left(\begin{array}{ccccccccccccc}
1\;\;&1\;\;&1\;\;&1\;\;&0\;\;&0\;\;&-1\;\;&0\;\;&0\;\;&0\;\;&0\;\;&-1\;\;&-1\;\;\\
1\;\;&1\;\;&1\;\;&1\;\;&0\;\;&0\;\;&0\;\;&-1\;\;&0\;\;&0\;\;&0\;\;&-1\;\;&-1\;\;\\
1\;\;&1\;\;&0\;\;&0\;\;&1\;\;&1\;\;&0\;\;&0\;\;&-1\;\;&0\;\;&0\;\;&-1\;\;&-1\;\;\\
1\;\;&1\;\;&0\;\;&0\;\;&1\;\;&1\;\;&0\;\;&0\;\;&0\;\;&-1\;\;&0\;\;&-1\;\;&-1\;\;\\
1\;\;&1\;\;&0\;\;&0\;\;&0\;\;&0\;\;&0\;\;&0\;\;&0\;\;&0\;\;&-1\;\;&0\;\;&0\;\;\\
\end{array}\right)
\;,\nonumber\\
&&{\vec{\vartheta}_{_{6B}}}^{\;T}=(\vartheta_{_{z_{_1}}},\cdots,\;\vartheta_{_{z_{_{18}}}})
\;,\nonumber\\
&&\mathbf{B_{_{6B}}}^{\;T}=(-a_{_1},\;\cdots,\;-a_{_6},\;b_{_1}-1,\;\cdots,\;b_{_7}-1)\;.
\label{GKZ6B18}
\end{eqnarray}
Here, $\mathbf{I_{_{13\times13}}}$ is a ${13\times13}$ unit matrix.

\section{GKZ hypergeometric system of 4-loop vacuum integrals with seven propagates\label{sec-seven}}

\subsection{4-loop vacuum diagram with seven propagates for U-type}

\begin{figure}[t]
\setlength{\unitlength}{0cm}
\centering
\vspace{-1cm}\hspace{1cm}
\includegraphics[width=9.0cm]{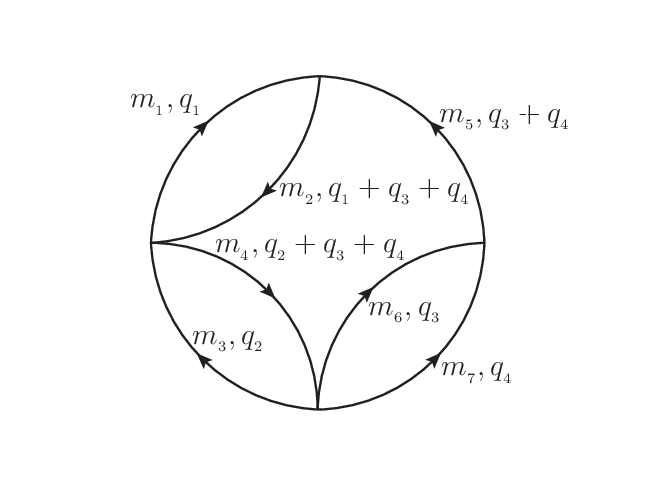}
\vspace{-1cm}
\caption[]{Four-loop vacuum diagram with seven propagators for U-type.}
\label{fig-4loop-4q}
\end{figure}

We picture a four-loop vacuum diagram with seven propagates in Fig.~\ref{fig-4loop-4q}. The diagram inside looks like the letter U, so we name it as U-type.
The U-type four-loop vacuum diagram has one master integral, which is
\begin{eqnarray}
&&U_{_{7A}}=\Big(\Lambda_{_{\rm RE}}^2\Big)^{8-2D}
\int{d^D {\bf q} \over(2\pi)^D}
{1 \over (q_{_1}^2-m_{_1}^2)[(q_{_1}+q_{_3}+q_{_4})^2-m_{_2}^2](q_{_2}^2-m_{_3}^2)}
\nonumber\\
&&\hspace{1.1cm}\times
{1\over
[(q_{_2}+q_{_3}+q_{_4})^2-m_{_4}^2][(q_{_3}+q_{_4})^2-m_{_5}^2](q_{_3}^2-m_{_6}^2)(q_{_4}^2-m_{_7}^2)}.
\label{GKZ7A0}
\end{eqnarray}
Integrating out ${\bf q}$, the Mellin-Barnes representation of the four-loop vacuum integral for U-type is written as
\begin{eqnarray}
&&U_{_{7A}}=
{-m_{_7}^2\over(2\pi i)^6(4\pi)^8}\Big({4\pi\Lambda_{_{\rm RE}}^2\over m_{_7}^2}\Big)^{8-2D}
\int_{-i\infty}^{+i\infty}d{\bf s}
\Big[\prod\limits_{i=1}^6\Big({m_{_i}^2\over m_{_7}^2}\Big)^{s_{_i}}\Gamma(-s_{_i})\Big]\nonumber\\
&&\hspace{1.0cm}\times
\Big[\prod\limits_{i=1}^4\Gamma({D\over2}-1-s_{_i})\Big]
\Gamma({D\over2}-1-s_{_6})\Gamma(6-{3D\over2}+\sum\limits_{i=1}^6s_{_i})
\Gamma(7-2D+\sum\limits_{i=1}^6s_{_i})\nonumber\\
&&\hspace{1.0cm}\times
{\Gamma({3D\over2}-5-\sum\limits_{i=1}^5s_{_i})
\Gamma(2-{D\over2}+s_{_1}+s_{_2})\Gamma(2-{D\over2}+s_{_3}+s_{_4})\Gamma(1+s_{_5})
\over\Gamma(5-D+\sum\limits_{i=1}^5s_{_i})\Gamma(D-2-s_{_1}-s_{_2})\Gamma(D-2-s_{_3}-s_{_4})}\;,
\label{GKZ7A2}
\end{eqnarray}
where ${\bf s}=(s_{_1},\;\cdots,\;s_{_6})$.

Then, we can derive one linear independent term of the integral:
\begin{eqnarray}
&&U_{_{7A}}\ni
{-m_{_7}^2\over(4\pi)^8}\Big({4\pi\Lambda_{_{\rm RE}}^2\over m_{_7}^2}\Big)^{8-2D}
{ \pi^4 \sin^2\pi D \over \sin^5\frac{\pi D}{2} \sin\frac{3\pi D}{2} }
T_{_{7A}}({\bf a},\;{\bf b}\;\Big|\;{\bf x})\;,
\label{GKZ7A5}
\end{eqnarray}
with
\begin{eqnarray}
T_{_{7A}}({\bf a},\;{\bf b}\;\Big|\;{\bf x})={\bf{\sum\limits_{{n}=0}^\infty}} A_{_{\bf n}} {\bf x^{n}},
\label{GKZ7A6}
\end{eqnarray}
where ${\bf n}=(n_{_1},\;\cdots,\;n_{_6})$, ${\bf x}=(x_{_1},\;\cdots,\;x_{_6})$, $x_{_i}={m_{_i}^2\over m_{_7}^2}$, ${\bf a}=(a_{_1},\;\cdots,a_{_7})$ and
${\bf b}=(b_{_1},\;\cdots,b_{_7})$ with
\begin{eqnarray}
&&a_{_1}=6-{3D\over2},\;a_{_2}=7-2D,\;a_{_3}=a_{_5}=2-{D\over2},\;
a_{_4}=a_{_6}=3-D,\;a_{_7}=1\;,
\nonumber\\
&&b_{_1}=b_{_2}=b_{_3}=b_{_4}=b_{_5}=2-{D\over2},\;b_{_6}=5-D,\;b_{_7}=6-{3D\over2}\:,
\label{GKZ7A6-1}
\end{eqnarray}
and the coefficient $A_{_{\bf n}}$ is
\begin{eqnarray}
\hspace{-1cm}&&A_{_{{\bf n}}}=
\Gamma(a_{_1}+\sum\limits_{i=1}^6n_{_i}) \Gamma(a_{_2}+\sum\limits_{i=1}^6n_{_i})
\Gamma(a_{_3}+\sum\limits_{i=1}^2n_{_i})
\nonumber\\
&&\hspace{1.0cm}\times
{\Gamma(a_{_4}+\sum\limits_{i=1}^2n_{_i})\Gamma(a_{_5}+\sum\limits_{i=3}^4n_{_i})
\Gamma(a_{_6}+\sum\limits_{i=3}^4n_{_i})\Gamma(a_{_7}+n_{_5})
\over  \Big[\prod\limits_{i=1}^6 n_{_i}!\Big]\Big[\prod\limits_{i=1}^4 \Gamma(b_{_i}+n_{_i})\Big]\Gamma(b_{_5}+n_{_6})
\Gamma(b_{_6}+\sum\limits_{i=1}^5n_{_i})\Gamma(b_{_7}+\sum\limits_{i=1}^5n_{_i})}.
\label{GKZ7A6-2}
\end{eqnarray}

We also define the auxiliary function
\begin{eqnarray}
&&\Phi_{_{7A}}({\bf a},\;{\bf b}\;\Big|\;{\bf x},\;{\bf u},\;{\bf v})={\bf u}^{\bf a}{\bf v}^{{\bf b}-{\bf e}_{_{7}}}
T_{_{7A}}({\bf a},\;{\bf b}\;\Big|\;{\bf x})\;,
\label{GKZ7A10}
\end{eqnarray}
with the intermediate variables ${\bf u}=(u_{_1},\cdots,u_{_7})$, ${\bf v}=(v_{_1},\cdots,v_{_7})$, ${\bf u}={\bf v}={\bf e}_{_{7}}=(1,\;1,\;1,\;1,\;1,\;1,\;1)$.
Then we can derive the GKZ hypergeometric system for the four-loop vacuum integral for U-type:
\begin{eqnarray}
&&\mathbf{A_{_{7A}}}\cdot\vec{\vartheta}_{_{7A}}\Phi_{_{7A}}=\mathbf{B_{_{7A}}}\Phi_{_{7A}}\;,
\label{GKZ7A17}
\end{eqnarray}
where
\begin{eqnarray}
&&\mathbf{A_{_{7A}}}=\left(\begin{array}{cc}
\mathbf{I_{_{14\times14}}}\;\;&\mathbf{A_{_{X7A}}}\;\;\\
\end{array}\right)
\;,\nonumber\\
&&\mathbf{A_{_{X7A}}}^{\;T}=\left(\begin{array}{cccccccccccccc}
1\;\;&1\;\;&1\;\;&1\;\;&0\;\;&0\;\;&0\;\;&-1\;\;&0\;\;&0\;\;&0\;\;&0\;\;&-1\;\;&-1\;\;\\
1\;\;&1\;\;&1\;\;&1\;\;&0\;\;&0\;\;&0\;\;&0\;\;&-1\;\;&0\;\;&0\;\;&0\;\;&-1\;\;&-1\;\;\\
1\;\;&1\;\;&0\;\;&0\;\;&1\;\;&1\;\;&0\;\;&0\;\;&0\;\;&-1\;\;&0\;\;&0\;\;&-1\;\;&-1\;\;\\
1\;\;&1\;\;&0\;\;&0\;\;&1\;\;&1\;\;&0\;\;&0\;\;&0\;\;&0\;\;&-1\;\;&0\;\;&-1\;\;&-1\;\;\\
1\;\;&1\;\;&0\;\;&0\;\;&0\;\;&0\;\;&1\;\;&0\;\;&0\;\;&0\;\;&0\;\;&0\;\;&-1\;\;&-1\;\;\\
1\;\;&1\;\;&0\;\;&0\;\;&0\;\;&0\;\;&0\;\;&0\;\;&0\;\;&0\;\;&0\;\;&-1\;\;&0\;\;&0\;\;\\
\end{array}\right)
\;,\nonumber\\
&&{\vec{\vartheta}_{_{7A}}}^{\;T}=(\vartheta_{_{z_{_1}}},\cdots,\;\vartheta_{_{z_{_{20}}}})
\;,\nonumber\\
&&\mathbf{B_{_{7A}}}^{\;T}=(-a_{_1},\;\cdots,\;-a_{_7},\;b_{_1}-1,\;\cdots,\;b_{_7}-1)\;.
\label{GKZ7A18}
\end{eqnarray}
Here, $\mathbf{I_{_{14\times14}}}$ is a ${14\times14}$ unit matrix.
Correspondingly the dual matrix $\mathbf{\tilde A_{_{7A}}}$ of $\mathbf{A_{_{7A}}}$ is
\begin{eqnarray}
&&\mathbf{\tilde A_{_{7A}}}=\left(\begin{array}{cc}
-\mathbf{A_{_{X7A}}}^{\;T}\;\;&\mathbf{I_{_{6\times6}}}\;\;\\
\end{array}\right),
\label{GKZ7A19}
\end{eqnarray}
where $\mathbf{I_{_{6\times6}}}$ is a ${6\times6}$ unit matrix.

\subsection{4-loop vacuum diagram with seven propagates for Z-type}

\begin{figure}[t]
\setlength{\unitlength}{0cm}
\centering
\vspace{-1cm}\hspace{1cm}
\includegraphics[width=9.0cm]{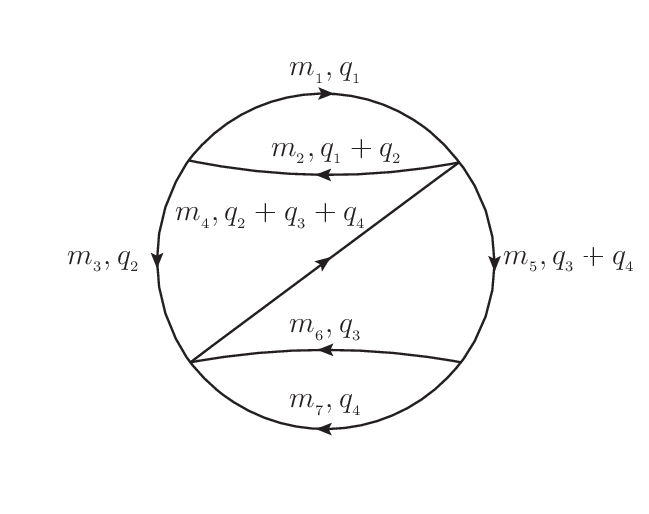}
\vspace{-1cm}
\caption[]{Four-loop vacuum diagram with seven propagators for Z-type.}
\label{fig-4loop-5q}
\end{figure}

In Fig.~\ref{fig-4loop-5q}, the diagram inside looks like the letter Z, so we name it as Z-type.
The master integral of the four-loop vacuum diagram with seven propagates for Z-type is
\begin{eqnarray}
&&U_{_{7B}}=\Big(\Lambda_{_{\rm RE}}^2\Big)^{8-2D}
\int{d^D {\bf q} \over(2\pi)^D}
{1 \over (q_{_1}^2-m_{_1}^2)[(q_{_1}+q_{_2})^2-m_{_2}^2](q_{_2}^2-m_{_3}^2)}
\nonumber\\
&&\hspace{1.1cm}\times
{1\over
[(q_{_2}+q_{_3}+q_{_4})^2-m_{_4}^2][(q_{_3}+q_{_4})^2-m_{_5}^2](q_{_3}^2-m_{_6}^2)(q_{_4}^2-m_{_7}^2)}.
\label{GKZ7B0}
\end{eqnarray}
One can give the Mellin-Barnes representation of the four-loop vacuum integral for Z-type:
\begin{eqnarray}
&&\hspace{-0.5cm}U_{_{7B}}=
{-m_{_7}^2\over(2\pi i)^6(4\pi)^8}\Big({4\pi\Lambda_{_{\rm RE}}^2\over m_{_7}^2}\Big)^{8-2D}
\int_{-i\infty}^{+i\infty}d{\bf s}
\Big[\prod\limits_{i=1}^6\Big({m_{_i}^2\over m_{_7}^2}\Big)^{s_{_i}}\Gamma(-s_{_i})\Big]
\Big[\prod\limits_{i=1}^2\Gamma({D\over2}-1-s_{_i})\Big]\nonumber\\
&&\hspace{-0.5cm}\hspace{1.0cm}\times
\Gamma({D\over2}-1-s_{_4})\Gamma({D\over2}-1-s_{_6})\Gamma(6-{3D\over2}+\sum\limits_{i=1}^6s_{_i})
\Gamma(7-2D+\sum\limits_{i=1}^6s_{_i})\Gamma(4-D+\sum\limits_{i=1}^4s_{_i})\nonumber\\
&&\hspace{-0.5cm}\hspace{1.0cm}\times
{\Gamma({3D\over2}-5-\sum\limits_{i=1}^5s_{_i})
\Gamma({D-3}-\sum\limits_{i=1}^3s_{_i})
\Gamma(2-{D\over2}+s_{_1}+s_{_2})\Gamma(1+s_{_3})\Gamma(1+s_{_5})
\over\Gamma(D-2-s_{_1}-s_{_2})\Gamma({3D\over2}-4-\sum\limits_{i=1}^4s_{_i})
\Gamma(3-{D\over2}+\sum\limits_{i=1}^3s_{_i})\Gamma(5-D+\sum\limits_{i=1}^5s_{_i})}\;.
\label{GKZ7B2}
\end{eqnarray}

Then, we can derive one linear independent term of the integral:
\begin{eqnarray}
&&U_{_{7B}}\ni
{m_{_7}^2\over(4\pi)^8}\Big({4\pi\Lambda_{_{\rm RE}}^2\over m_{_7}^2}\Big)^{8-2D}
{ \pi^4  \over \sin^4\frac{\pi D}{2}  }
T_{_{7B}}({\bf a},\;{\bf b}\;\Big|\;{\bf x})\;,
\label{GKZ7B5}
\end{eqnarray}
with
\begin{eqnarray}
T_{_{7B}}({\bf a},\;{\bf b}\;\Big|\;{\bf x})={\bf{\sum\limits_{{n}=0}^\infty}} A_{_{\bf n}} {\bf x^{n}},
\label{GKZ7B6}
\end{eqnarray}
where ${\bf a}=(a_{_1},\;\cdots,a_{_8})$ and
${\bf b}=(b_{_1},\;\cdots,b_{_8})$ with
\begin{eqnarray}
&&a_{_1}=6-{3D\over2},\;a_{_2}=7-2D,\;a_{_3}=5-{3D\over2},\;
a_{_4}=4-D,\;\nonumber\\
&&a_{_5}=2-{D\over2},\;
a_{_6}=3-D,\;a_{_7}=a_{_8}=1,\;b_{_1}=b_{_2}=b_{_3}=b_{_4}=2-{D\over2},\;
\nonumber\\
&&b_{_5}=3-{D\over2},\;
b_{_6}=4-D,\;b_{_7}=5-D,\;b_{_8}=6-{3D\over2}\:,
\label{GKZ7B6-1}
\end{eqnarray}
and the coefficient $A_{_{\bf n}}$ is
\begin{eqnarray}
\hspace{-1cm}&&A_{_{{\bf n}}}=
{\Gamma(a_{_1}+\sum\limits_{i=1}^6n_{_i}) \Gamma(a_{_2}+\sum\limits_{i=1}^6n_{_i})
\Gamma(a_{_3}+\sum\limits_{i=1}^4n_{_i})\Gamma(a_{_4}+\sum\limits_{i=1}^4n_{_i})
\over \Big[\prod\limits_{i=1}^6 n_{_i}!\Big] \Gamma(b_{_1}+n_{_1})\Gamma(b_{_2}+n_{_2})\Gamma(b_{_3}+n_{_4})\Gamma(b_{_4}+n_{_6})}
\nonumber\\
&&\hspace{1.0cm}\times
{\Gamma(a_{_5}+\sum\limits_{i=1}^2n_{_i})
\Gamma(a_{_6}+\sum\limits_{i=1}^2n_{_i})\Gamma(a_{_7}+n_{_3})\Gamma(a_{_8}+n_{_5})
\over \Gamma(b_{_5}+\sum\limits_{i=1}^3n_{_i})
\Gamma(b_{_6}+\sum\limits_{i=1}^3n_{_i})\Gamma(b_{_7}+\sum\limits_{i=1}^5n_{_i})
\Gamma(b_{_8}+\sum\limits_{i=1}^5n_{_i})}.
\label{GKZ7B6-2}
\end{eqnarray}

We also define the auxiliary function
\begin{eqnarray}
&&\Phi_{_{7B}}({\bf a},\;{\bf b}\;\Big|\;{\bf x},\;{\bf u},\;{\bf v})={\bf u}^{\bf a}{\bf v}^{{\bf b}-{\bf e}_{_{8}}}
T_{_{7B}}({\bf a},\;{\bf b}\;\Big|\;{\bf x})\;,
\label{GKZ7B10}
\end{eqnarray}
with the intermediate variables ${\bf u}=(u_{_1},\cdots,u_{_8})$, ${\bf v}=(v_{_1},\cdots,v_{_8})$, ${\bf u}={\bf v}={\bf e}_{_{8}}=(1,\;1,\;1,\;1,\;1,\;1,\;1,\;1)$.
Then we can give the GKZ hypergeometric system for the four-loop vacuum integral for Z-type:
\begin{eqnarray}
&&\mathbf{A_{_{7B}}}\cdot\vec{\vartheta}_{_{7B}}\Phi_{_{7B}}=\mathbf{B_{_{7B}}}\Phi_{_{7B}}\;,
\label{GKZ7B17}
\end{eqnarray}
where
\begin{eqnarray}
&&\mathbf{A_{_{7B}}}=\left(\begin{array}{cc}
\mathbf{I_{_{16\times16}}}\;\;&\mathbf{A_{_{X7B}}}\;\;\\
\end{array}\right)
\;,\nonumber\\
&&\mathbf{A_{_{X7B}}}^{\;T}=\left(\begin{array}{cccccccccccccccc}
1\;\;&1\;\;&1\;\;&1\;\;&1\;\;&1\;\;&0\;\;&0\;\;&-1\;\;&0\;\;&0\;\;&0\;\;&-1\;\;&-1\;\;&-1\;\;&-1\;\;\\
1\;\;&1\;\;&1\;\;&1\;\;&1\;\;&1\;\;&0\;\;&0\;\;&0\;\;&-1\;\;&0\;\;&0\;\;&-1\;\;&-1\;\;&-1\;\;&-1\;\;\\
1\;\;&1\;\;&1\;\;&1\;\;&0\;\;&0\;\;&1\;\;&0\;\;&0\;\;&0\;\;&0\;\;&0\;\;&-1\;\;&-1\;\;&-1\;\;&-1\;\;\\
1\;\;&1\;\;&1\;\;&1\;\;&0\;\;&0\;\;&0\;\;&0\;\;&0\;\;&0\;\;&-1\;\;&0\;\;&0\;\;&0\;\;&-1\;\;&-1\;\;\\
1\;\;&1\;\;&0\;\;&0\;\;&0\;\;&0\;\;&0\;\;&1\;\;&0\;\;&0\;\;&0\;\;&0\;\;&0\;\;&0\;\;&-1\;\;&-1\;\;\\
1\;\;&1\;\;&0\;\;&0\;\;&0\;\;&0\;\;&0\;\;&0\;\;&0\;\;&0\;\;&0\;\;&-1\;\;&0\;\;&0\;\;&0\;\;&0\;\;\\
\end{array}\right)
\;,\nonumber\\
&&{\vec{\vartheta}_{_{7B}}}^{\;T}=(\vartheta_{_{z_{_1}}},\cdots,\;\vartheta_{_{z_{_{22}}}})
\;,\nonumber\\
&&\mathbf{B_{_{7B}}}^{\;T}=(-a_{_1},\;\cdots,\;-a_{_8},\;b_{_1}-1,\;\cdots,\;b_{_8}-1)\;.
\label{GKZ7B18}
\end{eqnarray}
Here, $\mathbf{I_{_{16\times16}}}$ is a ${16\times16}$ unit matrix.

\section{The hypergeometric series solutions of 4-loop vacuum integrals\label{sec-solu}}

\subsection{The general case of 4-loop vacuum integral with five propagates}

In this subsection, we will show the hypergeometric series solutions of the GKZ hypergeometric system of 4-loop vacuum integral with five propagates with arbitrary masses.
To construct the hypergeometric series solutions of the GKZ hypergeometric system in Eq.~(\ref{4loop-20})
is equivalent to choose a set of the linear independent column vectors of the matrix in Eq.~(\ref{4loop-22})
which spans the dual space. We can denote the submatrix
composed of the first, third, fourth and fifth column vectors of the dual
matrix of Eq.~(\ref{4loop-22}) as $\mathbf{\tilde A}_{_{1345}}$, i.e.
\begin{eqnarray}
&&\mathbf{\tilde A}_{_{1345}}=\left(\begin{array}{cccc}
-1\;\;&1\;\;&0\;\;&0\;\;\\
-1\;\;&0\;\;&1\;\;&0\;\;\\
-1\;\;&0\;\;&0\;\;&1\;\;\\
-1\;\;&0\;\;&0\;\;&0\;\;
\end{array}\right)\;.
\label{4loop-S1}
\end{eqnarray}
Obviously $\det\mathbf{\tilde A}_{_{1345}}=1\neq0$, and
\begin{eqnarray}
&&\mathbf{B}_{_{1345}}=\mathbf{\tilde A}_{_{1345}}^{-1}\cdot\mathbf{\tilde A}_{_{5}}
\nonumber\\
&&\hspace{1.0cm}=\left(\begin{array}{cccccccccc}
1\;\;& 1\;\;& 0\;\;& 0\;\;& 0\;\;& -1\;\;& 0\;\;& 0\;\;& 0\;\;& -1\;\;\\
0\;\;& 0\;\;& 1\;\;& 0\;\;& 0\;\;& -1\;\;& 1\;\;& 0\;\;& 0\;\;& -1\;\;\\
0\;\;& 0\;\;& 0\;\;& 1\;\;& 0\;\;& -1\;\;& 0\;\;& 1\;\;& 0\;\;& -1\;\;\\
0\;\;& 0\;\;& 0\;\;& 0\;\;& 1\;\;& -1\;\;& 0\;\;& 0\;\;& 1\;\;& -1\;\;
\end{array}\right)\;.
\label{4loop-S2}
\end{eqnarray}
Taking 4 row vectors of the matrix $\mathbf{B}_{_{1345}}$ as the basis of integer lattice,
one can construct the hypergeometric series solutions in parameter space, through choosing the sets of column indices $I_{_i}\subset [1,\cdots,10]\;(i=1,\cdots,16)$ which are consistent with the basis of integer lattice  $\mathbf{B}_{_{1345}}$. In the following, we consider the denominators powers of the master integrals for 4-loop vacuum integral with five propagates are equal to 1. In another case, the relevant parameters can be replaced accordingly.

One can take the set of column indices
$I_{_1}=[2,6,\cdots,10]$, i.e. the implement $J_{_1}=[1,\cdots,10]\setminus I_{_1}=[1,3,4,5]$.
The choice on the set of indices implies the exponent numbers
$\alpha_{_1}=\alpha_{_3}=\alpha_{_4}=\alpha_{_5}=0$. Through Eq.~(\ref{4loop-26}), we have
\begin{eqnarray}
&&\alpha_{_2}=a_{_1}-a_{_2},\;\alpha_{_6}=\sum\limits_{i=1}^4 b_{_i}-a_{_1}-4,\;\alpha_{_7}=1-b_{_1},\;
\nonumber\\
&&\alpha_{_8}=1-b_{_2},\;\alpha_{_9}=1-b_{_3},\;\alpha_{_{10}}=\sum\limits_{i=1}^3 b_{_i}-a_{_1}-3\;.
\label{4loop-S3}
\end{eqnarray}
Combined with Eq.~(\ref{4loop-11-1}), one can have
\begin{eqnarray}
\alpha_{_2}=\frac{D}{2}-1,\;\alpha_{_6}=-\frac{D}{2},\;
\alpha_{_7}=\alpha_{_8}=\alpha_{_9}=\frac{D}{2}-1,\;\alpha_{_{10}}=-1\;.
\label{4loop-S4}
\end{eqnarray}
According the basis of integer lattice $\mathbf{B}_{_{1345}}$, the hypergeometric series solution with quadruple independent variables can be written as
\begin{eqnarray}
&&\Phi_{_{[1345]}}^{(1)}=\prod\limits_{i=1}^{10}z_{_i}^{\alpha_{_i}}\sum\limits_{{\bf n}=0}^\infty
{c_{_{[1345]}}^{(1)}({\bf n})}
\Big({z_{_1}z_{_2}\over z_{_6}z_{_{10}}}\Big)^{n_{_1}}
\Big({z_{_3}z_{_7}\over z_{_6}z_{_{10}}}\Big)^{n_{_2}}
\Big({z_{_4}z_{_8}\over z_{_6}z_{_{10}}}\Big)^{n_{_3}}
\Big({z_{_5}z_{_9}\over z_{_6}z_{_{10}}}\Big)^{n_{_4}}
\nonumber\\
&&\hspace{1.1cm}=
\prod\limits_{i=1}^{10}z_{_i}^{\alpha_{_i}}\sum\limits_{{\bf n}=0}^\infty
{c_{_{[1345]}}^{(1)}({\bf n})}
\Big({1\over y_{_4}}\Big)^{n_{_1}}
\Big({y_{_1}\over y_{_4}}\Big)^{n_{_2}}
\Big({y_{_2}\over y_{_4}}\Big)^{n_{_3}}\Big({y_{_3}\over y_{_4}}\Big)^{n_{_4}}\;,
\label{4loop-S5}
\end{eqnarray}
where the coefficient is
\begin{eqnarray}
&&c_{_{[1345]}}^{(1)}({\bf n})=\Big\{ \Big[\prod\limits_{i=1}^4 n_{_i}!\Big]\Gamma(1+\alpha_{_2}+n_{_1})\Gamma(1+\alpha_{_6}-\sum\limits_{i=1}^4 n_{_i})
\Gamma(1+\alpha_{_7}+n_{_2})\nonumber\\
&&\hspace{2.0cm}\times
\Gamma(1+\alpha_{_8}+n_{_3})\Gamma(1+\alpha_{_9}+n_{_4})
\Gamma(1+\alpha_{_{10}}-\sum\limits_{i=1}^4 n_{_i})
\Big\}^{-1}\;.
\label{4loop-S6}
\end{eqnarray}
Using the relation in Eq.~(\ref{Gamma}),
we can have
\begin{eqnarray}
&&\hspace{-1.2cm}c_{_{[1345]}}^{(1)}({\bf n})\propto
\frac{\Gamma(-\alpha_{_6}+\sum\limits_{i=1}^4 n_{_i}) \Gamma(-\alpha_{_{10}}+\sum\limits_{i=1}^4 n_{_i})}
{\Big[\prod\limits_{i=1}^4 n_{_i}!\Big]\Gamma(1+\alpha_{_2}+n_{_1})\Gamma(1+\alpha_{_7}+n_{_2})
\Gamma(1+\alpha_{_8}+n_{_3})\Gamma(1+\alpha_{_9}+n_{_4})},
\label{4loop-S8}
\end{eqnarray}
where we ignore the constant coefficient term ${\frac{\sin\pi\alpha_{_6} \sin\pi\alpha_{_{10}}}{\pi^2}}$.
And then, through Eq.~(\ref{4loop-S4}), the corresponding hypergeometric series solution is written as
\begin{eqnarray}
&&\Phi_{_{[1345]}}^{(1)}=
y_{_1}^{{D\over2}-1}y_{_2}^{{D\over2}-1}y_{_3}^{{D\over2}-1}y_{_4}^{-1}\sum\limits_{{\bf n}=0}^\infty
{c_{_{[1345]}}^{(1)}({\bf n})}f_{_{[1345]}}\;,
\nonumber\\
&&\;
f_{_{[1345]}}=\Big({1\over y_{_4}}\Big)^{n_{_1}}
\Big({y_{_1}\over y_{_4}}\Big)^{n_{_2}}
\Big({y_{_2}\over y_{_4}}\Big)^{n_{_3}}\Big({y_{_3}\over y_{_4}}\Big)^{n_{_4}}\;,
\label{4loop-S9}
\end{eqnarray}
with the coefficient is
\begin{eqnarray}
&&c_{_{[1345]}}^{(1)}({\bf n})=
\frac{\Gamma({D\over2}+\sum\limits_{i=1}^4 n_{_i}) \Gamma(1+\sum\limits_{i=1}^4 n_{_i})}
{\prod\limits_{i=1}^4 n_{_i}!\Gamma({D\over2}+n_{_i})}\;.
\label{4loop-S10}
\end{eqnarray}
Here, the convergent region
of the hypergeometric function $\Phi_{_{[1345]}}^{(1)}$ in Eq.~(\ref{4loop-S9}) is
\begin{eqnarray}
&&\Xi_{_{[1345]}}=\{(y_{_1},\;y_{_2},\;y_{_3},\;y_{_4})
\Big|1<|y_{_4}|,\;|y_{_1}|<|y_{_4}|,\;|y_{_2}|<|y_{_4}|,\;|y_{_3}|<|y_{_4}|\}\;,
\label{4loop-S11}
\end{eqnarray}
which shows that $\Phi_{_{[1345]}}^{(1)}$ is in neighborhood of regular singularity $\infty$.

According the basis of integer lattice $\mathbf{B}_{_{1345}}$, we can also obtain other fifteen hypergeometric solutions, which the expressions can be seen in Appendix~\ref{app1}. The sixteen hypergeometric series solutions $\Phi_{_{[1345]}}^{(i)}$ whose convergent region is  $\Xi_{_{[1345]}}$ can constitute a fundamental solution system.  The combination coefficients are determined by the value of the scalar integral of an ordinary point or some regular singularities.

Multiplying one of the row vectors of the matrix $\mathbf{B}_{_{1345}}$ by -1,
the induced integer matrix can be chosen as a basis of the integer lattice
space of certain hypergeometric series. Taking 4 row vectors of the following matrix as the
basis of integer lattice,
\begin{eqnarray}
&&\mathbf{B}_{_{\tilde{1}345}}={\rm diag}(-1,1,1,1)\cdot\mathbf{B}_{_{1345}}
\nonumber\\
&&\hspace{1.0cm}=\left(\begin{array}{cccccccccc}
-1\;\;& -1\;\;& 0\;\;& 0\;\;& 0\;\;& 1\;\;& 0\;\;& 0\;\;& 0\;\;& 1\;\;\\
0\;\;& 0\;\;& 1\;\;& 0\;\;& 0\;\;& -1\;\;& 1\;\;& 0\;\;& 0\;\;& -1\;\;\\
0\;\;& 0\;\;& 0\;\;& 1\;\;& 0\;\;& -1\;\;& 0\;\;& 1\;\;& 0\;\;& -1\;\;\\
0\;\;& 0\;\;& 0\;\;& 0\;\;& 1\;\;& -1\;\;& 0\;\;& 0\;\;& 1\;\;& -1\;\;
\end{array}\right)\;,
\label{4loop-S19}
\end{eqnarray}
one can obtain sixteen hypergeometric series solutions  $\Phi_{_{[\tilde{1}345]}}^{(i)}\:(i=1,\cdots,16)$  similarly, which the expressions can be seen in Appendix~\ref{app2}.
The convergent region of the functions $\Phi_{_{[\tilde{1}345]}}^{(i)}$ is
\begin{eqnarray}
&&\Xi_{_{[\tilde{1}345]}}=\{(y_{_1},\;y_{_2},\;y_{_3},\;y_{_4})
\Big||y_{_1}|<1,\;|y_{_2}|<1,\;|y_{_3}|<1,\;|y_{_4}|<1\}\;,
\label{4loop-S20}
\end{eqnarray}
which shows that $\Phi_{_{[\tilde{1}345]}}^{(i)}$ are in neighborhood of regular singularity $0$.

Taking 4 row vectors of the following matrix as the basis of integer lattice,
\begin{eqnarray}
&&\mathbf{B}_{_{1\tilde{3}45}}={\rm diag}(1,-1,1,1)\cdot\mathbf{B}_{_{1345}}
\nonumber\\
&&\hspace{1.0cm}=\left(\begin{array}{cccccccccc}
1\;\;& 1\;\;& 0\;\;& 0\;\;& 0\;\;& -1\;\;& 0\;\;& 0\;\;& 0\;\;& -1\;\;\\
0\;\;& 0\;\;& -1\;\;& 0\;\;& 0\;\;& 1\;\;& -1\;\;& 0\;\;& 0\;\;& 1\;\;\\
0\;\;& 0\;\;& 0\;\;& 1\;\;& 0\;\;& -1\;\;& 0\;\;& 1\;\;& 0\;\;& -1\;\;\\
0\;\;& 0\;\;& 0\;\;& 0\;\;& 1\;\;& -1\;\;& 0\;\;& 0\;\;& 1\;\;& -1\;\;
\end{array}\right)\;,
\label{4loop-S39}
\end{eqnarray}
one also obtains sixteen hypergeometric series solutions  $\Phi_{_{[1\tilde{3}45]}}^{(i)}\:(i=1,\cdots,16)$:
\begin{eqnarray}
&&\Phi_{_{[1\tilde{3}45]}}^{(i)}=\Phi_{_{[1345]}}^{(i)}(y_{_4}\leftrightarrow y_{_1})\;,
\label{4loop-S39-1}
\end{eqnarray}
which the expressions can be obtained by  interchanging between $y_{_4}$ and $y_{_1}$  in $\Phi_{_{[1345]}}^{(i)}$.
The convergent region of the functions $\Phi_{_{[1\tilde{3}45]}}^{(i)}$ is
\begin{eqnarray}
&&\Xi_{_{[1\tilde{3}45]}}=\{(y_{_1},\;y_{_2},\;y_{_3},\;y_{_4})
\Big|1<|y_{_1}|,\;|y_{_4}|<|y_{_1}|,\;|y_{_2}|<|y_{_1}|,\;|y_{_3}|<|y_{_1}|\}\;,
\label{4loop-S40}
\end{eqnarray}
which shows that $\Phi_{_{[1\tilde{3}45]}}^{(i)}$ are in neighborhood of regular singularity $\infty$.

Taking 4 row vectors of the following matrix as the basis of integer lattice,
\begin{eqnarray}
&&\mathbf{B}_{_{13\tilde{4}5}}={\rm diag}(1,1,-1,1)\cdot\mathbf{B}_{_{1345}}
\nonumber\\
&&\hspace{1.0cm}=\left(\begin{array}{cccccccccc}
1\;\;& 1\;\;& 0\;\;& 0\;\;& 0\;\;& -1\;\;& 0\;\;& 0\;\;& 0\;\;& -1\;\;\\
0\;\;& 0\;\;& 1\;\;& 0\;\;& 0\;\;& -1\;\;& 1\;\;& 0\;\;& 0\;\;& -1\;\;\\
0\;\;& 0\;\;& 0\;\;& -1\;\;& 0\;\;& 1\;\;& 0\;\;& -1\;\;& 0\;\;& 1\;\;\\
0\;\;& 0\;\;& 0\;\;& 0\;\;& 1\;\;& -1\;\;& 0\;\;& 0\;\;& 1\;\;& -1\;\;
\end{array}\right)\;,
\label{4loop-S49}
\end{eqnarray}
one can obtain sixteen hypergeometric series solutions  $\Phi_{_{[13\tilde{4}5]}}^{(i)}\:(i=1,\cdots,16)$:
\begin{eqnarray}
&&\Phi_{_{[13\tilde{4}5]}}^{(i)}=\Phi_{_{[1345]}}^{(i)}(y_{_4}\leftrightarrow y_{_2})\;,
\label{4loop-S49-1}
\end{eqnarray}
which the expressions can be obtained by  interchanging between $y_{_4}$ and $y_{_2}$  in $\Phi_{_{[1345]}}^{(i)}$.
The convergent region of the functions $\Phi_{_{[13\tilde{4}5]}}^{(i)}$ is
\begin{eqnarray}
&&\Xi_{_{[13\tilde{4}5]}}=\{(y_{_1},\;y_{_2},\;y_{_3},\;y_{_4})
\Big|1<|y_{_2}|,\;|y_{_1}|<|y_{_2}|,\;|y_{_4}|<|y_{_2}|,\;|y_{_3}|<|y_{_2}|\}\;,
\label{4loop-S50}
\end{eqnarray}
which shows that $\Phi_{_{[13\tilde{4}5]}}^{(i)}$ are in neighborhood of regular singularity $\infty$.

Taking 4 row vectors of the following matrix as the basis of integer lattice,
\begin{eqnarray}
&&\mathbf{B}_{_{134\tilde{5}}}={\rm diag}(1,1,1,-1)\cdot\mathbf{B}_{_{1345}}
\nonumber\\
&&\hspace{1.0cm}=\left(\begin{array}{cccccccccc}
1\;\;& 1\;\;& 0\;\;& 0\;\;& 0\;\;& -1\;\;& 0\;\;& 0\;\;& 0\;\;& -1\;\;\\
0\;\;& 0\;\;& 1\;\;& 0\;\;& 0\;\;& -1\;\;& 1\;\;& 0\;\;& 0\;\;& -1\;\;\\
0\;\;& 0\;\;& 0\;\;& 1\;\;& 0\;\;& -1\;\;& 0\;\;& 1\;\;& 0\;\;& -1\;\;\\
0\;\;& 0\;\;& 0\;\;& 0\;\;& -1\;\;& 1\;\;& 0\;\;& 0\;\;& -1\;\;& 1\;\;
\end{array}\right)\;,
\label{4loop-S59}
\end{eqnarray}
one can obtain sixteen hypergeometric series solutions  $\Phi_{_{[134\tilde{5}]}}^{(i)}\:(i=1,\cdots,16)$:
\begin{eqnarray}
&&\Phi_{_{[134\tilde{5}]}}^{(i)}=\Phi_{_{[1345]}}^{(i)}(y_{_4}\leftrightarrow y_{_3})\;,
\label{4loop-S59-1}
\end{eqnarray}
which the expressions can be obtained by  interchanging between $y_{_4}$ and $y_{_3}$  in $\Phi_{_{[1345]}}^{(i)}$.
The convergent region of the functions $\Phi_{_{[134\tilde{5}]}}^{(i)}$ is
\begin{eqnarray}
&&\Xi_{_{[134\tilde{5}]}}=\{(y_{_1},\;y_{_2},\;y_{_3},\;y_{_4})
\Big|1<|y_{_3}|,\;|y_{_1}|<|y_{_3}|,\;|y_{_2}|<|y_{_3}|,\;|y_{_4}|<|y_{_3}|\}\;,
\label{4loop-S60}
\end{eqnarray}
which shows that $\Phi_{_{[134\tilde{5}]}}^{(i)}$ are in neighborhood of regular singularity $\infty$.

\subsection{The special case of 4-loop vacuum integral with five propagates}

In order to elucidate how to obtain the analytical expression of 4-loop vacuum integral clearly, we assume the two nonzero virtual mass of 4-loop vacuum integral with five propagates.
The special case of 4-loop vacuum integral can be expressed as a linear combination of those corresponding functionally independent Gauss functions.

Through Sec. \ref{sec-five}, the GKZ hypergeometric system in this special case ($m_{_1}\neq 0,\;m_{_5}\neq 0,\;m_{_2}=m_{_3}=m_{_4}=0$ with $\nu_{_i}=1$) is simplified as
\begin{eqnarray}
&&\mathbf{A_{_{51}}}\cdot\vec{\vartheta}_{_{51}}\Phi_{_{51}}=\mathbf{B_{_{5}}}\Phi_{_{51}}\;,
\label{GKZ34a}
\end{eqnarray}
where the vector of Euler operators is defined as
\begin{eqnarray}
&&\vec{\vartheta}_{_{51}}^{\;T}=(\vartheta_{_{z_{_1}}},\cdots,\;\vartheta_{_{z_{_{7}}}})\;,
\label{GKZ35a}
\end{eqnarray}
and the matrix $\mathbf{A}_{_{51}}$ is  obtained through deleting the 8th,
9th, and 10th columns of the matrix $\mathbf{A_{_{5}}}$:
\begin{eqnarray}
&&\mathbf{A}_{_{51}}=\left(\begin{array}{ccccccc}
1\;\;&0\;\;&0\;\;&0\;\;&0\;\;&0\;\;&1\;\\
0\;\;&1\;\;&0\;\;&0\;\;&0\;\;&0\;\;&1\;\\
0\;\;&0\;\;&1\;\;&0\;\;&0\;\;&0\;\;&-1\;\\
0\;\;&0\;\;&0\;\;&1\;\;&0\;\;&0\;\;&0\;\\
0\;\;&0\;\;&0\;\;&0\;\;&1\;\;&0\;\;&0\;\\
0\;\;&0\;\;&0\;\;&0\;\;&0\;\;&1\;\;&0\;\\
\end{array}\right)\;.
\label{GKZ36a}
\end{eqnarray}
Through Eq.~(\ref{GKZ34a}), we can have the relations
\begin{eqnarray}
&&\alpha_{_1}+\alpha_{_{7}}=-a_{_1}\;,
\quad \alpha_{_2}+\alpha_{_{7}}=-a_{_2}\;,
\quad \alpha_{_3}-\alpha_{_{7}}=b_{_1}-1\;,
\label{GKZ36a-1}
\end{eqnarray}
with $a_{_1}=4-{3D\over2}$, $a_{_2}=5-2D$, $b_{_1}=2-{D\over2}$, and the other $\alpha_{_i}$ are zero.
The dual matrix $\mathbf{\tilde A}_{_{51}}$ of $\mathbf{A}_{_{51}}$ is
\begin{eqnarray}
&&\mathbf{\tilde A}_{_{51}}=\left(\begin{array}{ccccccc}
-1\;\;&-1\;\;&1\;\;&0\;\;&0\;\;&0\;\;&1\;\;
\end{array}\right)\;.
\label{GKZ37a}
\end{eqnarray}

The integer sublattice $\mathbf{B_{_{51}}}$ is determined by the dual matrix $\mathbf{\tilde A}_{_{51}}$ with $\mathbf{B_{_{51}}}=\mathbf{\tilde A}_{_{51}}$.
The integer sublattice $\mathbf{B_{_{51}}}$  implies that the system of fundamental solutions
is composed by two linear independent hypergeometric functions.
One can take the set of column indices
$I_{_1}=[1,\cdots,6]$, which implies the exponent numbers $\alpha_{_{7}}=0$.
And then, the hypergeometric series solution is written as
\begin{eqnarray}
&&\Phi_{_{[51]}}^{(1)}(y_{_1})=\;_{_2}F_{_1}
\left(\left.\begin{array}{cc}4-{3D\over2},\;&5-2D\\
\;\;&2-{D\over2}\end{array}\right|y_{_1}\right)\;,
\label{GKZ38a1}
\end{eqnarray}
with $y_{_1}=x_{_1}=m_{_1}^2/m_{_5}^2$, and $_{_2}F_{_1}$ is Gauss function:
\begin{eqnarray}
&&_{_2}F_{_1}
\left(\left.\begin{array}{cc}a,\;&b\\
\;\;&c\end{array}\right|x\right)
=\;\sum\limits_{n=0}^\infty{(a)_n(b)_n
\over n!(c)_n}x^n\;,
\label{GKZ38a1-1}
\end{eqnarray}
with $(a)_n=\Gamma(a+n)/\Gamma(a)$. One also can take the set of column indices
$I_{_2}=[1,2,4,\cdots,7]$, which implies the exponent numbers $\alpha_{_{3}}=0$.
And then, the another hypergeometric series solution is written as
\begin{eqnarray}
&&\Phi_{_{[51]}}^{(2)}(y_{_1})=
(y_{_1})^{D/2-1}\;_{_2}F_{_1}
\left(\left.\begin{array}{cc}3-D,\;&4-{3D\over2}\\
\;\;&{D\over2}\end{array}\right|y_{_1}\right)\;.
\label{GKZ38a2}
\end{eqnarray}
Here, the convergent region of the functions $\Phi_{_{[51]}}^{(1,2)}(y_{_1})$ is $|y_{_1}|<1$.
In the region $|y_{_1}|<1$, the integral is a linear combination of two fundamental solutions:
\begin{eqnarray}
&&\Phi_{_{51}}(y_{_1})=C_{_{[51]}}^{(1)}\Phi_{_{[51]}}^{(1)}(y_{_1})
+C_{_{[51]}}^{(2)}\Phi_{_{[51]}}^{(2)}(y_{_1})\;.
\label{GKZ38a3}
\end{eqnarray}

Multiplying one of the row vectors of the integer matrix $\mathbf{B}_{_{51}}$ by -1,
the induced integer matrix can be chosen as a basis of the integer lattice
space.
And the corresponding system of
fundamental solutions is similarly composed by two Gauss functions:
\begin{eqnarray}
&&\Phi_{_{[51]}}^{(3)}(y_{_1})=
(y_{_1})^{{3D\over2}-4}\;_{_2}F_{_1}
\left(\left.\begin{array}{cc}4-{3D\over2},\;&3-D\\
\;\;&{D\over2}\end{array}\right|{1\over y_{_1}}\right)\;,
\nonumber\\
&&\Phi_{_{[51]}}^{(4)}(y_{_1})=
(y_{_1})^{2D-5}\;_{_2}F_{_1}
\left(\left.\begin{array}{cc}5-2D,\;&4-{3D\over2}\\
\;\;&2-{D\over2}\end{array}\right|{1\over y_{_1}}\right)\;,
\label{GKZ39a}
\end{eqnarray}
which the convergent region is $|y_{_1}|>1$.
Correspondingly in the region $|y_{_1}|>1$, the integral is a linear combination of two fundamental solutions:
\begin{eqnarray}
&&\Phi_{_{51}}(y_{_1})=C_{_{[51]}}^{(3)}\Phi_{_{[51]}}^{(3)}(y_{_1})
+C_{_{[51]}}^{(4)}\Phi_{_{[51]}}^{(4)}(y_{_1})\;.
\label{GKZ39a1}
\end{eqnarray}

As $m_{_1}^2\ll m_{_5}^2,\;m_{_2}=m_{_3}=m_{_4}=0$,
\begin{eqnarray}
&&I_{_1}=\Big(\Lambda_{_{\rm RE}}^2\Big)^{8-2D}\int{d^D{\bf{q}}\over(2\pi)^D}
{1\over(q_{_1}^2-m_{_1}^2)q_{_2}^2q_{_3}^2(q_{_1}+q_{_2}+q_{_3}+q_{_4})^2
(q_{_3}^2-m_{_5}^2)}
\nonumber\\
&&\hspace{0.5cm}=I_{_{1,0}}+\cdots\;,
\label{GKZ40a}
\end{eqnarray}
where
\begin{eqnarray}
&&I_{_{1,0}}=\Big(\Lambda_{_{\rm RE}}^2\Big)^{8-2D}\int{d^D{\bf{q}}\over(2\pi)^D}
{1\over q_{_1}^2q_{_2}^2q_{_3}^2(q_{_1}+q_{_2}+q_{_3}+q_{_4})^2
(q_{_3}^2-m_{_5}^2)}
\nonumber\\
&&\hspace{0.7cm}=
{-m_{_5}^6\over(4\pi)^8}\Big({4\pi\Lambda_{_{\rm RE}}^2\over m_{_5}^2}\Big)^{8-{2D}}
\Gamma(4-{3D\over2})\Gamma(5-{2D})\;.
\label{GKZ41a}
\end{eqnarray}
This result indicates
\begin{eqnarray}
&&C_{_{[51]}}^{(1)}={-m_{_5}^6\over(4\pi)^8}\Big({4\pi\Lambda_{_{\rm RE}}^2\over m_{_5}^2}\Big)^{8-{2D}}
\Gamma(4-{3D\over2})\Gamma(5-{2D})\;.
\label{GKZ41a-1}
\end{eqnarray}

As $m_{_1}^2\gg m_{_5}^2$ and $\;m_{_2}=m_{_3}=m_{_4}=0$, $I_{_1}=I_{_{1,\infty}}+\cdots\;$, where
\begin{eqnarray}
&&I_{_{1,\infty}}=\Big(\Lambda_{_{\rm RE}}^2\Big)^{8-2D}\int{d^D{\bf{q}}\over(2\pi)^D}
{1\over(q_{_1}^2-m_{_1}^2)q_{_2}^2q_{_3}^2(q_{_1}+q_{_2}+q_{_3}+q_{_4})^2
q_{_3}^2}
\nonumber\\
&&\hspace{0.8cm}=
{-m_{_1}^6\over(4\pi)^8}\Big({4\pi\Lambda_{_{\rm RE}}^2\over m_{_1}^2}\Big)^{8-{2D}}
\Gamma(4-{3D\over2})\Gamma(5-{2D})\;.
\label{GKZ44a}
\end{eqnarray}
This result indicates
\begin{eqnarray}
&&C_{_{[51]}}^{(4)}={-m_{_5}^6\over(4\pi)^8}\Big({4\pi\Lambda_{_{\rm RE}}^2\over m_{_5}^2}\Big)^{8-{2D}}
\Gamma(4-{3D\over2})\Gamma(5-{2D})\;.
\label{GKZ44a-1}
\end{eqnarray}

Actually, the Mellin-Barnes representation
of the integral in this case can be obtained as
\begin{eqnarray}
&&\hspace{-0.5cm}U_{_5}=
{-m_{_5}^6\over(4\pi)^8}\Big({4\pi\Lambda_{_{\rm RE}}^2\over m_{_5}^2}\Big)^{8-{2D}}
{1\over2\pi i}\int_{-i\infty}^{+i\infty}ds_{_1}
\Big({m_{_1}^2\over m_{_5}^2}\Big)^{s_{_1}}\Gamma(-s_{_1})
\nonumber\\
&&\hspace{0.5cm}\times
\Gamma({D\over2}-1-s_{_1})\Gamma(4-{3D\over2}+s_{_1})
\Gamma(5-{{2 D}}+s_{_1})\;.
\label{GKZ44a-4}
\end{eqnarray}
The residue of simple pole of $\Gamma(-s_{_1})$ provides $C_{_{[51]}}^{(1)}\Phi_{_{[51]}}^{(1)}(y_{_1})$,
that of simple pole of $\Gamma(D/2-1-s_{_1})$ provides $C_{_{[51]}}^{(2)}\Phi_{_{[51]}}^{(2)}(y_{_1})$,
that of simple pole of $\Gamma(4-{3D\over2}+s_{_1})$ provides $C_{_{[51]}}^{(3)}\Phi_{_{[51]}}^{(3)}(y_{_1})$,
and that of simple pole of $\Gamma(5-2 D+s_{_1})$ provides $C_{_{[51]}}^{(4)}\Phi_{_{[51]}}^{(4)}(y_{_1})$,
respectively.
And the residue of the simple pole of $\Gamma(D/2-1-s_{_1})$ and $\Gamma(4-{3D\over2}+s_{_1})$  can induce
\begin{eqnarray}
&&C_{_{[51]}}^{(2)}=C_{_{[51]}}^{(3)}={-m_{_5}^6\over(4\pi)^8}\Big({4\pi\Lambda_{_{\rm RE}}^2\over m_{_5}^2}\Big)^{8-{2D}}
\Gamma(4-{3D\over2})\Gamma(3-{D})\Gamma(1-{D\over2})\;.
\label{GKZ44a-5}
\end{eqnarray}

A conclusion for the other two nonzero virtual mass for the four-loop vacuum integral with five propagates, such as $m_{_2}\neq0$, $m_{_5}\neq0$, $m_{_1}=m_{_3}=m_{_4}=0$ is analogous
to that of $m_{_1}\neq0$, $m_{_5}\neq0$, $m_{_2}=m_{_3}=m_{_4}=0$.

\subsection{4-loop vacuum integrals with six or seven propagates}

In our previous works~\cite{Feng2020,GKZ-2loop}, we present GKZ hypergeometric systems of some one-loop and two-loop Feynman integrals, which show the algorithm and the obvious  hypergeometric series solutions for the one-loop and two-loop Feynman integrals. Recently, the authors of the Refs.~\cite{Ananthanarayan2021,Ananthanarayan2022GKZ} also give publicly available computer packages MBConicHulls~\cite{Ananthanarayan2021} and FeynGKZ~\cite{Ananthanarayan2022GKZ} to compute Feynman integrals in terms of hypergeometric functions, which are meaningful to improve computing efficiency. Through the package FeynGKZ~\cite{Ananthanarayan2022GKZ}, the authors give the examples of some one-loop and two-loop Feynman integrals, which are tested analytically by our previous work~\cite{Feng2020,GKZ-2loop}, as well as numerically using the package FIESTA~\cite{Smirnov-FIESTA5}.

Here, we also evaluate the four-loop vacuum integrals with six or seven propagates using FeynGKZ~\cite{Ananthanarayan2022GKZ}, which can be seen in the supplementary material. Note that FeynGKZ can't evaluate the four-loop vacuum integrals with five propagates through Mellin-Barnes representations and Miller's transformation, due that this option is still at an experimental stage in the package. In the supplementary material, we can see that the  GKZ hypergeometric systems of the four-loop vacuum integrals with six or seven propagates are in agree with our above results. Using FeynGKZ, we also give some hypergeometric series solutions for the four-loop vacuum integrals with six or seven propagates. The series solutions from GKZ hypergeometric systems for special case with the two nonzero virtual masses and the three nonzero virtual masses are also showed in the supplementary material, and tested numerically using FIESTA~\cite{Smirnov-FIESTA5}. One can see that the computing time using the hypergeometric series solutions is about ${\mathcal{O}}(10^{-5})$ times that using FIESTA, which evaluate quickly.

Being different from the package FeynGKZ, we try to give more fundamental solution systems in different convergent regions. By analyzing the fundamental solution systems, we hope to give the generalized Gauss relations for the hypergeometric series solutions~\cite{Gauss}. Finally, we would give the analytic solutions of Feynman integrals and write the corresponding computer package including the analytic functions, what we are going to do in next work.

Note that, except above five topologies with 6 master integrals, the four-loop vacuum integrals still have five topologies with 7 master integrals, which are one topology with seven propagates,  two topologies with eight propagates, and  two topologies with nine propagates \cite{R4loop4}. Due that they have complex mathematical structure, the five topologies of the four-loop vacuum integrals can't obtain the GKZ hypergeometric systems, through Mellin-Barnes representations and Miller's transformation. In next work, we will embed the general four-loop vacuum Feynman integrals into the subvarieties of Grassmannian manifold~\cite{Grassmannians}, to explore more possibilities of the general four-loop vacuum Feynman integrals.

\section{Conclusions\label{sec-con}}

Using Mellin-Barnes representation and Miller's transformation, we present GKZ hypergeometric systems of 4-loop vacuum integrals with arbitrary masses.
The dimension of the GKZ hypergeometric system equals the number of independent
dimensionless ratios among the virtual mass squared.
In the neighborhoods of origin and infinity, one can obtain the hypergeometric series
solutions of 4-loop vacuum integrals  through GKZ hypergeometric systems.
The linear independent hypergeometric series solutions whose convergent regions have
non-empty intersection can constitute a fundamental solution system in a proper subset
of the whole parameter space.
In certain convergent region, the four-loop vacuum integrals can be formulated as a linear combination of the fundamental solution system.
The combination coefficients can be determined by
the integral at some ordinary points or regular singularities, the Mellin-Barnes representation of the integral, or some mathematical methods.

Here, we can obtain the hypergeometric solutions of 4-loop vacuum integrals in the neighborhoods of origin including infinity, using GKZ hypergeometric systems on general manifold. In order to derive the fundamental solution system
in neighborhoods of all possible regular singularities, next we will embed the
vacuum integrals in corresponding Grassmannian manifold~\cite{Grassmannians}. And we also try to give the analytic solutions of the integrals in general form through generalized Gauss relations~\cite{Gauss}.

\begin{acknowledgments}

The work has been supported by the National Natural Science Foundation of China (NNSFC) with Grants No. 12075074, No. 12235008, Hebei Natural Science Foundation with Grants No. A2022201017, No. A2023201041, Natural Science Foundation of Guangxi Autonomous Region with Grant No. 2022GXNSFDA035068, and the youth top-notch talent support program of the Hebei Province.

\end{acknowledgments}

\appendix

\section{The calculation of the integral $I_{q}$\label{app-Iq}}

For the integral $I_{q}$, firstly we can integrate out $q_{_1}$:
\begin{eqnarray}
I_{q_{_1}}&&\equiv \int{d^Dq_{_1}\over(2\pi)^D}{1\over(q_{_1}^2)^{\nu_{_1}+s_{_1}}
[(q_{_1}+q_{_2}+q_{_3}+q_{_4})^2]^{\nu_{_4}+s_{_4}}
}\nonumber\\
&&={\Gamma(\nu_{_1}+\nu_{_4}+s_{_1}+s_{_4})\over\Gamma(\nu_{_1}+s_{_1})\Gamma(\nu_{_4}+s_{_4})}
\int_0^1dx(1-x)^{\nu_{_1}-1+s_{_1}}x^{\nu_{_4}-1+s_{_4}}\nonumber\\
&&\hspace{0.4cm}\times
\int{d^Dq_{_1}\over(2\pi)^D}
{1\over[q_{_1}^2+x(1-x)(q_{_2}+q_{_3}+q_{_4})^2]^{\nu_{_1}+\nu_{_4}+s_{_1}+s_{_4}}}\:.
\label{Iq-3}
\end{eqnarray}
Using the well-known integral
\begin{eqnarray}
\int{d^Dq\over(2\pi)^D}{1\over[q^2+{\triangle}]^n}
={i(-)^{D/2}\Gamma(n-{D\over2})\over (4\pi)^{D/2}\Gamma(n)}{\Big( {1\over {\triangle}} \Big)}^{n-D/2}\:,
\label{integral}
\end{eqnarray}
one can have
\begin{eqnarray}
I_{q_{_1}}&&=
{i(-)^{D/2}\Gamma(\nu_{_1}+\nu_{_4}-{D\over2}+s_{_1}+s_{_4})\over(4\pi)^{D/2}\Gamma(\nu_{_1}+s_{_1})\Gamma(\nu_{_4}+s_{_4})}
{1\over [(q_{_2}+q_{_3}+q_{_4})^2]^{\nu_{_1}+\nu_{_4}-{D\over2}+s_{_1}+s_{_4}}}
\nonumber\\
&&\hspace{0.4cm}\times
\int_0^1dx x^{D/2-1-\nu_{_1}-s_{_1}}(1-x)^{D/2-1-\nu_{_4}-s_{_4}}\:.
\label{Iq-3-1}
\end{eqnarray}
Through Beta function
\begin{eqnarray}
B(m,n)=\int_0^1dx\: x^{m-1}(1-x)^{n-1} ={ \Gamma(m)\Gamma(n)\over \Gamma(m+n)}\:,
\label{Beta}
\end{eqnarray}
we can have
\begin{eqnarray}
I_{{q_{_1}}}&&=
{i(-)^{D/2}\Gamma(\nu_{_1}+\nu_{_4}-{D\over2}+s_{_1}+s_{_4})\Gamma({D\over2}-\nu_{_1}-s_{_1})\Gamma({D\over2}-\nu_{_4}-s_{_4})
\over(4\pi)^{D/2}\Gamma(\nu_{_1}+s_{_1})\Gamma(\nu_{_4}+s_{_4})\Gamma(D-\nu_{_1}-\nu_{_4}-s_{_1}-s_{_4})}
\nonumber\\
&&\hspace{0.4cm}\times{1\over [(q_{_2}+q_{_3}+q_{_4})^2]^{\nu_{_1}+\nu_{_4}-{D\over2}+s_{_1}+s_{_4}}}\:.
\label{Iq-4}
\end{eqnarray}

Similarly, we can integrate out $q_{_2}$:
\begin{eqnarray}
I_{q_{_2}}&&\equiv \int{d^Dq_{_2}\over(2\pi)^D}{1\over(q_{_2}^2)^{\nu_{_2}+s_{_2}}
[(q_{_2}+q_{_3}+q_{_4})^2]^{\nu_{_1}+\nu_{_4}-{D\over2}+s_{_1}+s_{_4}}}
\nonumber\\
&&=
{i(-)^{D/2}\Gamma(\nu_{_1}+\nu_{_2}+\nu_{_4}-D+s_{_1}+s_{_2}+s_{_4})\Gamma({D\over2}-\nu_{_2}-s_{_2})
\Gamma(D-\nu_{_1}-\nu_{_4}-s_{_1}-s_{_4})
\over(4\pi)^{D/2}\Gamma(\nu_{_2}+s_{_2})\Gamma(\nu_{_1}+\nu_{_4}-{D\over2}+s_{_1}+s_{_4})
\Gamma({3D\over2}-\nu_{_1}-\nu_{_2}-\nu_{_4}-s_{_1}-s_{_2}-s_{_4})}
\nonumber\\
&&\hspace{0.4cm}\times{1\over [(q_{_3}+q_{_4})^2]^{\nu_{_1}+\nu_{_2}+\nu_{_4}-D+s_{_1}+s_{_2}+s_{_4}}}\:.
\label{Iq-5}
\end{eqnarray}

And then, we integrate out $q_{_3}$:
\begin{eqnarray}
I_{q_{_3}}&&\equiv \int{d^Dq_{_3}\over(2\pi)^D}{1\over(q_{_3}^2)^{\nu_{_3}+s_{_3}}
[(q_{_3}+q_{_4})^2]^{\nu_{_1}+\nu_{_2}+\nu_{_4}-D+s_{_1}+s_{_2}+s_{_4}}}
\nonumber\\
&&=
{i(-)^{D/2}\Gamma(\sum\limits_{i=1}^4 \nu_{_i}-{3D\over2}+\sum\limits_{i=1}^4s_{_i})
\Gamma({D\over2}-\nu_{_3}-s_{_3})\Gamma({3D\over2}-\nu_{_1}-\nu_{_2}-\nu_{_4}-s_{_1}-s_{_2}-s_{_4})
\over(4\pi)^{D/2}\Gamma(\nu_{_3}+s_{_3})\Gamma(\nu_{_1}+\nu_{_2}+\nu_{_4}-D+s_{_1}+s_{_2}+s_{_4})\Gamma(2D-\sum\limits_{i=1}^4 \nu_{_i}-\sum\limits_{i=1}^4s_{_i})}
\nonumber\\
&&\hspace{0.4cm}\times{1\over (q_{_4}^2)^{\sum\limits_{i=1}^4 \nu_{_i}-{3D\over2}+\sum\limits_{i=1}^4s_{_i}}}\:.
\label{Iq-5-1}
\end{eqnarray}

Last, through the parameterization
\begin{eqnarray}
{1\over A^m B^n}={\Gamma(m+n) \over \Gamma(m)\Gamma(n)} \int_0^1{dx} {x^{m-1}(1-x)^{n-1}\over [xA+(1-x)B]^{m+n}}\:,
\end{eqnarray}
we integrate out $q_{_4}$:
\begin{eqnarray}
I_{q_{_4}}&&\equiv\int{d^Dq_{_4}\over(2\pi)^D}
{1\over(q_{_4}^2-m_{_5}^2)^{\nu_{_5}}(q_{_4}^2)^{\sum\limits_{i=1}^4 \nu_{_i}-{3D\over2}+\sum\limits_{i=1}^4s_{_i}}}
\nonumber\\
&&\hspace{-0.0cm}={{i} \over (4\pi)^{D/2}\Gamma(\nu_{_5})} (-)^{\sum\limits_{i=1}^5 \nu_{_i}-{3D\over2}+\sum\limits_{i=1}^4s_{_i}} \Big( {1\over m_{_5}^2}\Big)^{\sum\limits_{i=1}^5 \nu_{_i}-2D+\sum\limits_{i=1}^4s_{_i}}
\nonumber\\
&&\hspace{0.4cm}\times{\Gamma(\sum\limits_{i=1}^5 \nu_{_i}-2D+\sum\limits_{i=1}^4s_{_i}) \Gamma(2D-\sum\limits_{i=1}^4 \nu_{_i}-\sum\limits_{i=1}^4s_{_i}) }\:.
\label{Iq-6}
\end{eqnarray}

\section{The hypergeometric series solutions of the integer lattice ${\bf B}_{_{1345}}$\label{app1}}

According the basis of integer lattice  $\mathbf{B}_{_{1345}}$,
one can construct other GKZ hypergeometric series solutions below.
\begin{itemize}
\item $J_{_2}=[1,3,4,9]$:
\begin{eqnarray}
&&\Phi_{_{[1345]}}^{(2)}=
y_{_1}^{{D\over2}-1}y_{_2}^{{D\over2}-1}y_{_4}^{{D\over2}-2}\sum\limits_{{\bf n}=0}^\infty
{c_{_{[1345]}}^{(2)}({\bf n})}f_{_{[1345]}}\;,\nonumber\\
&&c_{_{[1345]}}^{(2)}({\bf n})=
\frac{\Gamma(1+\sum\limits_{i=1}^4 n_{_i}) \Gamma(2-{D\over2}+\sum\limits_{i=1}^4 n_{_i})}
{\Big[\prod\limits_{i=1}^4 n_{_i}!\Big]\Gamma({D\over2}+n_{_1})\Gamma({D\over2}+n_{_2})
\Gamma({D\over2}+n_{_3})\Gamma(2-{D\over2}+n_{_4})}\;.
\label{4loop-S11-2}
\end{eqnarray}

\item $J_{_3}=[1,3,5,8]$:
\begin{eqnarray}
&&\Phi_{_{[1345]}}^{(3)}=
y_{_1}^{{D\over2}-1}y_{_3}^{{D\over2}-1}y_{_4}^{{D\over2}-2}\sum\limits_{{\bf n}=0}^\infty
{c_{_{[1345]}}^{(3)}({\bf n})}f_{_{[1345]}}\;,\nonumber\\
&&c_{_{[1345]}}^{(3)}({\bf n})=
\frac{\Gamma(1+\sum\limits_{i=1}^4 n_{_i}) \Gamma(2-{D\over2}+\sum\limits_{i=1}^4 n_{_i})}
{\Big[\prod\limits_{i=1}^4 n_{_i}!\Big]\Gamma({D\over2}+n_{_1})\Gamma({D\over2}+n_{_2})
\Gamma(2-{D\over2}+n_{_3})\Gamma({D\over2}+n_{_4})}\;.
\label{4loop-S11-3}
\end{eqnarray}

\item $J_{_4}=[1,3,8,9]$:
\begin{eqnarray}
&&\Phi_{_{[1345]}}^{(4)}=
y_{_1}^{{D\over2}-1}y_{_4}^{{D}-3}\sum\limits_{{\bf n}=0}^\infty
{c_{_{[1345]}}^{(4)}({\bf n})}f_{_{[1345]}}\;,\nonumber\\
&&c_{_{[1345]}}^{(4)}({\bf n})=
\frac{\Gamma(2-{D\over2}+\sum\limits_{i=1}^4 n_{_i}) \Gamma(3-D+\sum\limits_{i=1}^4 n_{_i})}
{\Big[\prod\limits_{i=1}^4 n_{_i}!\Big]\Gamma({D\over2}+n_{_1})\Gamma({D\over2}+n_{_2})
\Gamma(2-{D\over2}+n_{_3})\Gamma(2-{D\over2}+n_{_4})}\;.
\label{4loop-S11-4}
\end{eqnarray}

\item $J_{_5}=[1,4,5,7]$:
\begin{eqnarray}
&&\Phi_{_{[1345]}}^{(5)}=
y_{_2}^{{D\over2}-1}y_{_3}^{{D\over2}-1}y_{_4}^{{D\over2}-2}\sum\limits_{{\bf n}=0}^\infty
{c_{_{[1345]}}^{(5)}({\bf n})}f_{_{[1345]}}\;,\nonumber\\
&&c_{_{[1345]}}^{(5)}({\bf n})=
\frac{\Gamma(1+\sum\limits_{i=1}^4 n_{_i}) \Gamma(2-{D\over2}+\sum\limits_{i=1}^4 n_{_i})}
{\Big[\prod\limits_{i=1}^4 n_{_i}!\Big]\Gamma({D\over2}+n_{_1})\Gamma(2-{D\over2}+n_{_2})
\Gamma({D\over2}+n_{_3})\Gamma({D\over2}+n_{_4})}\;.
\label{4loop-S11-5}
\end{eqnarray}

\item $J_{_6}=[1,4,7,9]$:
\begin{eqnarray}
&&\Phi_{_{[1345]}}^{(6)}=
y_{_2}^{{D\over2}-1}y_{_4}^{{D}-3}\sum\limits_{{\bf n}=0}^\infty
{c_{_{[1345]}}^{(6)}({\bf n})}f_{_{[1345]}}\;,\nonumber\\
&&c_{_{[1345]}}^{(6)}({\bf n})=
\frac{\Gamma(2-{D\over2}+\sum\limits_{i=1}^4 n_{_i}) \Gamma(3-{D}+\sum\limits_{i=1}^4 n_{_i})}
{\Big[\prod\limits_{i=1}^4 n_{_i}!\Big]\Gamma({D\over2}+n_{_1})\Gamma(2-{D\over2}+n_{_2})
\Gamma({D\over2}+n_{_3})\Gamma(2-{D\over2}+n_{_4})}\;.
\label{4loop-S11-6}
\end{eqnarray}

\item $J_{_7}=[1,5,7,8]$:
\begin{eqnarray}
&&\Phi_{_{[1345]}}^{(7)}=
y_{_3}^{{D\over2}-1}y_{_4}^{{D}-3}\sum\limits_{{\bf n}=0}^\infty
{c_{_{[1345]}}^{(7)}({\bf n})}f_{_{[1345]}}\;,\nonumber\\
&&c_{_{[1345]}}^{(7)}({\bf n})=
\frac{\Gamma(2-{D\over2}+\sum\limits_{i=1}^4 n_{_i}) \Gamma(3-{D}+\sum\limits_{i=1}^4 n_{_i})}
{\Big[\prod\limits_{i=1}^4 n_{_i}!\Big]\Gamma({D\over2}+n_{_1})\Gamma(2-{D\over2}+n_{_2})
\Gamma(2-{D\over2}+n_{_3})\Gamma({D\over2}+n_{_4})}\;.
\label{4loop-S11-7}
\end{eqnarray}

\item $J_{_8}=[1,7,8,9]$:
\begin{eqnarray}
&&\hspace{-0.8cm}\Phi_{_{[1345]}}^{(8)}=
y_{_4}^{{3D\over2}-4}\sum\limits_{{\bf n}=0}^\infty
{c_{_{[1345]}}^{(8)}({\bf n})}f_{_{[1345]}}\;,\nonumber\\
&&\hspace{-0.8cm}c_{_{[1345]}}^{(8)}({\bf n})=
\frac{\Gamma(3-D+\sum\limits_{i=1}^4 n_{_i}) \Gamma( 4-{3D\over2}+\sum\limits_{i=1}^4 n_{_i})}
{\Big[\prod\limits_{i=1}^4 n_{_i}!\Big]\Gamma({D\over2}+n_{_1})\Gamma(2-{D\over2}+n_{_2})
\Gamma(2-{D\over2}+n_{_3})\Gamma(2-{D\over2}+n_{_4})}\;.
\label{4loop-S11-8}
\end{eqnarray}

\item $J_{_9}=[2,3,4,5]$:
\begin{eqnarray}
&&\Phi_{_{[1345]}}^{(9)}=
y_{_1}^{{D\over2}-1}y_{_2}^{{D\over2}-1}y_{_3}^{{D\over2}-1}y_{_4}^{{D\over2}-2}\sum\limits_{{\bf n}=0}^\infty
{c_{_{[1345]}}^{(9)}({\bf n})}f_{_{[1345]}}\;,\nonumber\\
&&c_{_{[1345]}}^{(9)}({\bf n})=
\frac{\Gamma(1+\sum\limits_{i=1}^4 n_{_i}) \Gamma(2-{D\over2}+\sum\limits_{i=1}^4 n_{_i})}
{\Big[\prod\limits_{i=1}^4 n_{_i}!\Big]\Gamma(2-{D\over2}+n_{_1})\Gamma({D\over2}+n_{_2})
\Gamma({D\over2}+n_{_3})\Gamma({D\over2}+n_{_4})}\;.
\label{4loop-S11-9}
\end{eqnarray}

\item $J_{_{10}}=I_{_{10}}=[2,3,4,9]$:
\begin{eqnarray}
&&\Phi_{_{[1345]}}^{(10)}=
y_{_1}^{{D\over2}-1}y_{_2}^{{D\over2}-1}y_{_4}^{{D}-3}\sum\limits_{{\bf n}=0}^\infty
{c_{_{[1345]}}^{(10)}({\bf n})}f_{_{[1345]}}\;,\nonumber\\
&&c_{_{[1345]}}^{(10)}({\bf n})=
\frac{\Gamma(2-{D\over2}+\sum\limits_{i=1}^4 n_{_i}) \Gamma(3-{D}+\sum\limits_{i=1}^4 n_{_i})}
{\Big[\prod\limits_{i=1}^4 n_{_i}!\Big]\Gamma(2-{D\over2}+n_{_1})\Gamma({D\over2}+n_{_2})
\Gamma({D\over2}+n_{_3})\Gamma(2-{D\over2}+n_{_4})}\;.
\label{4loop-S11-10}
\end{eqnarray}

\item $J_{_{11}}=[2,3,5,8]$:
\begin{eqnarray}
&&\Phi_{_{[1345]}}^{(11)}=
y_{_1}^{{D\over2}-1}y_{_3}^{{D\over2}-1}y_{_4}^{{D}-3}\sum\limits_{{\bf n}=0}^\infty
{c_{_{[1345]}}^{(11)}({\bf n})}f_{_{[1345]}}\;,\nonumber\\
&&c_{_{[1345]}}^{(11)}({\bf n})=
\frac{\Gamma(2-{D\over2}+\sum\limits_{i=1}^4 n_{_i}) \Gamma(3-{D}+\sum\limits_{i=1}^4 n_{_i})}
{\Big[\prod\limits_{i=1}^4 n_{_i}!\Big]\Gamma(2-{D\over2}+n_{_1})\Gamma({D\over2}+n_{_2})
\Gamma(2-{D\over2}+n_{_3})\Gamma({D\over2}+n_{_4})}\;.
\label{4loop-S11-11}
\end{eqnarray}

\item $J_{_{12}}=[2,3,8,9]$:
\begin{eqnarray}
&&\Phi_{_{[1345]}}^{(12)}=
y_{_1}^{{D\over2}-1}y_{_4}^{{3D\over2}-4}\sum\limits_{{\bf n}=0}^\infty
{c_{_{[1345]}}^{(12)}({\bf n})}f_{_{[1345]}}\;,\nonumber\\
&&c_{_{[1345]}}^{(12)}({\bf n})=
\frac{\Gamma(3-D+\sum\limits_{i=1}^4 n_{_i}) \Gamma(4-{3D\over2}+\sum\limits_{i=1}^4 n_{_i})}
{\Big[\prod\limits_{i=1}^4 n_{_i}!\Big]\Gamma(2-{D\over2}+n_{_1})\Gamma({D\over2}+n_{_2})
\Gamma(2-{D\over2}+n_{_3})\Gamma(2-{D\over2}+n_{_4})}\;.
\label{4loop-S11-12}
\end{eqnarray}

\item $J_{_{13}}=[2,4,5,7]$:
\begin{eqnarray}
&&\hspace{-0.5cm}\Phi_{_{[1345]}}^{(13)}=
y_{_2}^{{D\over2}-1}y_{_3}^{{D\over2}-1}y_{_4}^{{D}-3}\sum\limits_{{\bf n}=0}^\infty
{c_{_{[1345]}}^{(13)}({\bf n})}f_{_{[1345]}}\;,\nonumber\\
&&\hspace{-0.5cm}c_{_{[1345]}}^{(13)}({\bf n})=
\frac{\Gamma(2-{D\over2}+\sum\limits_{i=1}^4 n_{_i}) \Gamma(3-{D}+\sum\limits_{i=1}^4 n_{_i})}
{\Big[\prod\limits_{i=1}^4 n_{_i}!\Big]\Gamma(2-{D\over2}+n_{_1})\Gamma(2-{D\over2}+n_{_2})
\Gamma({D\over2}+n_{_3})\Gamma({D\over2}+n_{_4})}\;.
\label{4loop-S11-13}
\end{eqnarray}

\item $J_{_{14}}=[2,4,7,9]$:
\begin{eqnarray}
&&\hspace{-1.0cm}\Phi_{_{[1345]}}^{(14)}=
y_{_2}^{{D\over2}-1}y_{_4}^{{3D\over2}-4}\sum\limits_{{\bf n}=0}^\infty
{c_{_{[1345]}}^{(14)}({\bf n})}f_{_{[1345]}}\;,\nonumber\\
&&\hspace{-1.0cm}c_{_{[1345]}}^{(14)}({\bf n})=
\frac{\Gamma(3-{D}+\sum\limits_{i=1}^4 n_{_i}) \Gamma(4-{3D\over2}+\sum\limits_{i=1}^4 n_{_i})}
{\Big[\prod\limits_{i=1}^4 n_{_i}!\Big]\Gamma(2-{D\over2}+n_{_1})\Gamma(2-{D\over2}+n_{_2})
\Gamma({D\over2}+n_{_3})\Gamma(2-{D\over2}+n_{_4})}\;.
\label{4loop-S11-14}
\end{eqnarray}

\item $J_{_{15}}=[2,5,7,8]$:
\begin{eqnarray}
&&\hspace{-1.0cm}\Phi_{_{[1345]}}^{(15)}=
y_{_3}^{{D\over2}-1}y_{_4}^{{3D\over2}-4}\sum\limits_{{\bf n}=0}^\infty
{c_{_{[1345]}}^{(15)}({\bf n})}f_{_{[1345]}}\;,\nonumber\\
&&\hspace{-1.0cm}c_{_{[1345]}}^{(15)}({\bf n})=
\frac{\Gamma(3-{D}+\sum\limits_{i=1}^4 n_{_i}) \Gamma(4-{3D\over2}+\sum\limits_{i=1}^4 n_{_i})}
{\Big[\prod\limits_{i=1}^4 n_{_i}!\Big]\Gamma(2-{D\over2}+n_{_1})\Gamma(2-{D\over2}+n_{_2})
\Gamma(2-{D\over2}+n_{_3})\Gamma({D\over2}+n_{_4})}\;.
\label{4loop-S11-15}
\end{eqnarray}

\item $J_{_{16}}=[2,7,8,9]$:
\begin{eqnarray}
&&\hspace{-1.8cm}\Phi_{_{[1345]}}^{(16)}=
y_{_4}^{{2D}-5}\sum\limits_{{\bf n}=0}^\infty
{c_{_{[1345]}}^{(16)}({\bf n})}f_{_{[1345]}}\;,\nonumber\\
&&\hspace{-1.8cm}c_{_{[1345]}}^{(16)}({\bf n})=
\frac{\Gamma(4-{3D\over2}+\sum\limits_{i=1}^4 n_{_i}) \Gamma(5-2D+\sum\limits_{i=1}^4 n_{_i})}
{\prod\limits_{i=1}^4 n_{_i}!\Gamma(2-{D\over2}+n_{_i})}\;.
\label{4loop-S11-16}
\end{eqnarray}

\end{itemize}

\section{The hypergeometric series solutions of the integer lattice ${\bf B}_{_{\tilde{1}345}}$\label{app2}}

According the basis of integer lattice  $\mathbf{B}_{_{\tilde{1}345}}$,
one can construct GKZ hypergeometric series solutions below.
\begin{itemize}
\item $J_{_{1}}=[3,4,5,6]$:
\begin{eqnarray}
&&\Phi_{_{[\tilde{1}345]}}^{(1)}=
y_{_1}^{{D\over2}-1}y_{_2}^{{D\over2}-1}y_{_3}^{{D\over2}-1}y_{_4}^{{D\over2}-1}\sum\limits_{{\bf n}=0}^\infty
{c_{_{[\tilde{1}345]}}^{(1)}({\bf n})}f_{_{[\tilde{1}345]}}\;,
\nonumber\\
&&\;
f_{_{[\tilde{1}345]}}=\Big({y_{_4}}\Big)^{n_{_1}}
\Big({y_{_1}}\Big)^{n_{_2}}
\Big({y_{_2}}\Big)^{n_{_3}}\Big(y_{_3}\Big)^{n_{_4}}\;,
\label{4loop-S21-0}
\end{eqnarray}
where the coefficient is
\begin{eqnarray}
&&c_{_{[\tilde{1}345]}}^{(1)}({\bf n})=
\frac{\Gamma({D\over2}+\sum\limits_{i=1}^4 n_{_i}) \Gamma(1+\sum\limits_{i=1}^4 n_{_i})}
{\prod\limits_{i=1}^4 n_{_i}!\Gamma({D\over2}+n_{_i})}\;.
\label{4loop-S21-1}
\end{eqnarray}

\item $J_{_{2}}=[3,4,5,10]$:
\begin{eqnarray}
&&\Phi_{_{[\tilde{1}345]}}^{(2)}=
y_{_1}^{{D\over2}-1}y_{_2}^{{D\over2}-1}y_{_3}^{{D\over2}-1}\sum\limits_{{\bf n}=0}^\infty
{c_{_{[\tilde{1}345]}}^{(2)}({\bf n})}f_{_{[\tilde{1}345]}}\;,\nonumber\\
&&c_{_{[\tilde{1}345]}}^{(2)}({\bf n})=
\frac{\Gamma(1+\sum\limits_{i=1}^4 n_{_i}) \Gamma(2-{D\over2}+\sum\limits_{i=1}^4 n_{_i})}
{\Big[\prod\limits_{i=1}^4 n_{_i}!\Big]\Gamma(2-{D\over2}+n_{_1})\Gamma({D\over2}+n_{_2})
\Gamma({D\over2}+n_{_3})\Gamma({D\over2}+n_{_4})}\;.
\label{4loop-S21-2}
\end{eqnarray}

\item $J_{_{3}}=[3,4,6,9]$:
\begin{eqnarray}
&&\Phi_{_{[\tilde{1}345]}}^{(3)}=
y_{_1}^{{D\over2}-1}y_{_2}^{{D\over2}-1}y_{_4}^{{D\over2}-1}\sum\limits_{{\bf n}=0}^\infty
{c_{_{[\tilde{1}345]}}^{(3)}({\bf n})}f_{_{[\tilde{1}345]}}\;,\nonumber\\
&&c_{_{[\tilde{1}345]}}^{(3)}({\bf n})=
\frac{\Gamma(1+\sum\limits_{i=1}^4 n_{_i}) \Gamma(2-{D\over2}+\sum\limits_{i=1}^4 n_{_i})}
{\Big[\prod\limits_{i=1}^4 n_{_i}!\Big]\Gamma({D\over2}+n_{_1})\Gamma({D\over2}+n_{_2})
\Gamma({D\over2}+n_{_3})\Gamma(2-{D\over2}+n_{_4})}\;.
\label{4loop-S21-3}
\end{eqnarray}

\item $J_{_{4}}=[3,4,9,10]$:
\begin{eqnarray}
&&\Phi_{_{[\tilde{1}345]}}^{(4)}=
y_{_1}^{{D\over2}-1}y_{_2}^{{D\over2}-1}\sum\limits_{{\bf n}=0}^\infty
{c_{_{[\tilde{1}345]}}^{(4)}({\bf n})}f_{_{[\tilde{1}345]}}\;,\nonumber\\
&&c_{_{[\tilde{1}345]}}^{(4)}({\bf n})=
\frac{\Gamma(2-{D\over2}+\sum\limits_{i=1}^4 n_{_i}) \Gamma(3-{D}+\sum\limits_{i=1}^4 n_{_i})}
{\Big[\prod\limits_{i=1}^4 n_{_i}!\Big]\Gamma(2-{D\over2}+n_{_1})\Gamma({D\over2}+n_{_2})
\Gamma({D\over2}+n_{_3})\Gamma(2-{D\over2}+n_{_4})}\;.
\label{4loop-S21-4}
\end{eqnarray}

\item $J_{_{5}}=[3,5,6,8]$:
\begin{eqnarray}
&&\Phi_{_{[\tilde{1}345]}}^{(5)}=
y_{_1}^{{D\over2}-1}y_{_3}^{{D\over2}-1}y_{_4}^{{D\over2}-1}\sum\limits_{{\bf n}=0}^\infty
{c_{_{[\tilde{1}345]}}^{(5)}({\bf n})}f_{_{[\tilde{1}345]}}\;,
\nonumber\\
&&c_{_{[\tilde{1}345]}}^{(5)}({\bf n})=
\frac{\Gamma(1+\sum\limits_{i=1}^4 n_{_i}) \Gamma(2-{D\over2}+\sum\limits_{i=1}^4 n_{_i})}
{\Big[\prod\limits_{i=1}^4 n_{_i}!\Big]\Gamma({D\over2}+n_{_1})\Gamma({D\over2}+n_{_2})
\Gamma(2-{D\over2}+n_{_3})\Gamma({D\over2}+n_{_4})}\;.
\label{4loop-S21-5}
\end{eqnarray}

\item $J_{_{6}}=[3,5,8,10]$:
\begin{eqnarray}
&&\Phi_{_{[\tilde{1}345]}}^{(6)}=
y_{_1}^{{D\over2}-1}y_{_3}^{{D\over2}-1}\sum\limits_{{\bf n}=0}^\infty
{c_{_{[\tilde{1}345]}}^{(6)}({\bf n})}f_{_{[\tilde{1}345]}}\;,\nonumber\\
&&c_{_{[\tilde{1}345]}}^{(6)}({\bf n})=
\frac{\Gamma(2-{D\over2}+\sum\limits_{i=1}^4 n_{_i}) \Gamma(3-{D}+\sum\limits_{i=1}^4 n_{_i})}
{\Big[\prod\limits_{i=1}^4 n_{_i}!\Big]\Gamma(2-{D\over2}+n_{_1})\Gamma({D\over2}+n_{_2})
\Gamma(2-{D\over2}+n_{_3})\Gamma({D\over2}+n_{_4})}\;.
\label{4loop-S21-6}
\end{eqnarray}

\item $J_{_{7}}=[3,6,8,9]$:
\begin{eqnarray}
&&\Phi_{_{[\tilde{1}345]}}^{(7)}=
y_{_1}^{{D\over2}-1}y_{_4}^{{D\over2}-1}\sum\limits_{{\bf n}=0}^\infty
{c_{_{[\tilde{1}345]}}^{(7)}({\bf n})}f_{_{[\tilde{1}345]}}\;,\nonumber\\
&&c_{_{[\tilde{1}345]}}^{(7)}({\bf n})=
\frac{\Gamma(2-{D\over2}+\sum\limits_{i=1}^4 n_{_i}) \Gamma(3-{D}+\sum\limits_{i=1}^4 n_{_i})}
{\Big[\prod\limits_{i=1}^4 n_{_i}!\Big]\Gamma({D\over2}+n_{_1})\Gamma({D\over2}+n_{_2})
\Gamma(2-{D\over2}+n_{_3})\Gamma(2-{D\over2}+n_{_4})}\;.
\label{4loop-S21-7}
\end{eqnarray}

\item $J_{_{8}}=[3,8,9,10]$:
\begin{eqnarray}
&&\hspace{-1.0cm}\Phi_{_{[\tilde{1}345]}}^{(8)}=
y_{_1}^{{D\over2}-1}\sum\limits_{{\bf n}=0}^\infty
{c_{_{[\tilde{1}345]}}^{(8)}({\bf n})}f_{_{[\tilde{1}345]}}\;,\nonumber\\
&&\hspace{-1.0cm}c_{_{[\tilde{1}345]}}^{(8)}({\bf n})=
\frac{\Gamma(3-{D}+\sum\limits_{i=1}^4 n_{_i}) \Gamma(4-{3D\over2}+\sum\limits_{i=1}^4 n_{_i})}
{\Big[\prod\limits_{i=1}^4 n_{_i}!\Big]\Gamma(2-{D\over2}+n_{_1})\Gamma({D\over2}+n_{_2})
\Gamma(2-{D\over2}+n_{_3})\Gamma(2-{D\over2}+n_{_4})}\;.
\label{4loop-S21-8}
\end{eqnarray}

\item $J_{_{9}}=[4,5,6,7]$:
\begin{eqnarray}
&&\Phi_{_{[\tilde{1}345]}}^{(9)}=
y_{_2}^{{D\over2}-1}y_{_3}^{{D\over2}-1}y_{_4}^{{D\over2}-1}\sum\limits_{{\bf n}=0}^\infty
{c_{_{[\tilde{1}345]}}^{(9)}({\bf n})}f_{_{[\tilde{1}345]}}\;,
\nonumber\\
&&c_{_{[\tilde{1}345]}}^{(9)}({\bf n})=
\frac{\Gamma(1+\sum\limits_{i=1}^4 n_{_i}) \Gamma(2-{D\over2}+\sum\limits_{i=1}^4 n_{_i})}
{\Big[\prod\limits_{i=1}^4 n_{_i}!\Big]\Gamma({D\over2}+n_{_1})\Gamma(2-{D\over2}+n_{_2})
\Gamma({D\over2}+n_{_3})\Gamma({D\over2}+n_{_4})}\;.
\label{4loop-S21-9}
\end{eqnarray}

\item $J_{_{10}}=[4,5,7,10]$:
\begin{eqnarray}
&&\hspace{-0.8cm}\Phi_{_{[\tilde{1}345]}}^{(10)}=
y_{_2}^{{D\over2}-1}y_{_3}^{{D\over2}-1}\sum\limits_{{\bf n}=0}^\infty
{c_{_{[\tilde{1}345]}}^{(10)}({\bf n})}f_{_{[\tilde{1}345]}}\;,\nonumber\\
&&\hspace{-0.8cm}c_{_{[\tilde{1}345]}}^{(10)}({\bf n})=
\frac{\Gamma(2-{D\over2}+\sum\limits_{i=1}^4 n_{_i}) \Gamma(3-{D}+\sum\limits_{i=1}^4 n_{_i})}
{\Big[\prod\limits_{i=1}^4 n_{_i}!\Big]\Gamma(2-{D\over2}+n_{_1})\Gamma(2-{D\over2}+n_{_2})
\Gamma({D\over2}+n_{_3})\Gamma({D\over2}+n_{_4})}\;.
\label{4loop-S21-10}
\end{eqnarray}

\item $J_{_{11}}=[4,6,7,9]$:
\begin{eqnarray}
&&\hspace{-0.8cm}\Phi_{_{[\tilde{1}345]}}^{(11)}=
y_{_2}^{{D\over2}-1}y_{_4}^{{D\over2}-1}\sum\limits_{{\bf n}=0}^\infty
{c_{_{[\tilde{1}345]}}^{(11)}({\bf n})}f_{_{[\tilde{1}345]}}\;,\nonumber\\
&&\hspace{-0.8cm}c_{_{[\tilde{1}345]}}^{(11)}({\bf n})=
\frac{\Gamma(2-{D\over2}+\sum\limits_{i=1}^4 n_{_i}) \Gamma(3-{D}+\sum\limits_{i=1}^4 n_{_i})}
{\Big[\prod\limits_{i=1}^4 n_{_i}!\Big]\Gamma({D\over2}+n_{_1})\Gamma(2-{D\over2}+n_{_2})
\Gamma({D\over2}+n_{_3})\Gamma(2-{D\over2}+n_{_4})}\;.
\label{4loop-S21-11}
\end{eqnarray}

\item $J_{_{12}}=[4,7,9,10]$:
\begin{eqnarray}
&&\hspace{-1.cm}\Phi_{_{[\tilde{1}345]}}^{(12)}=
y_{_2}^{{D\over2}-1}\sum\limits_{{\bf n}=0}^\infty
{c_{_{[\tilde{1}345]}}^{(12)}({\bf n})}f_{_{[\tilde{1}345]}}\;,\nonumber\\
&&\hspace{-1.cm}c_{_{[\tilde{1}345]}}^{(12)}({\bf n})=
\frac{\Gamma(3-{D}+\sum\limits_{i=1}^4 n_{_i}) \Gamma(4-{3D\over2}+\sum\limits_{i=1}^4 n_{_i})}
{\Big[\prod\limits_{i=1}^4 n_{_i}!\Big]\Gamma(2-{D\over2}+n_{_1})\Gamma(2-{D\over2}+n_{_2})
\Gamma({D\over2}+n_{_3})\Gamma(2-{D\over2}+n_{_4})}\;.
\label{4loop-S21-12}
\end{eqnarray}

\item $J_{_{13}}=[5,6,7,8]$:
\begin{eqnarray}
&&\hspace{-0.8cm}\Phi_{_{[\tilde{1}345]}}^{(13)}=
y_{_3}^{{D\over2}-1}y_{_4}^{{D\over2}-1}\sum\limits_{{\bf n}=0}^\infty
{c_{_{[\tilde{1}345]}}^{(13)}({\bf n})}f_{_{[\tilde{1}345]}}\;,
\nonumber\\
&&\hspace{-0.8cm}c_{_{[\tilde{1}345]}}^{(13)}({\bf n})=
\frac{\Gamma(2-{D\over2}+\sum\limits_{i=1}^4 n_{_i}) \Gamma(3-{D}+\sum\limits_{i=1}^4 n_{_i})}
{\Big[\prod\limits_{i=1}^4 n_{_i}!\Big]\Gamma({D\over2}+n_{_1})\Gamma(2-{D\over2}+n_{_2})
\Gamma(2-{D\over2}+n_{_3})\Gamma({D\over2}+n_{_4})}\;.
\label{4loop-S21-13}
\end{eqnarray}

\item $J_{_{14}}=[5,7,8,10]$:
\begin{eqnarray}
&&\hspace{-1cm}\Phi_{_{[\tilde{1}345]}}^{(14)}=
y_{_3}^{{D\over2}-1}\sum\limits_{{\bf n}=0}^\infty
{c_{_{[\tilde{1}345]}}^{(14)}({\bf n})}f_{_{[\tilde{1}345]}}\;,\nonumber\\
&&\hspace{-1cm}c_{_{[\tilde{1}345]}}^{(14)}({\bf n})=
\frac{\Gamma(3-{D}+\sum\limits_{i=1}^4 n_{_i}) \Gamma(4-{3D\over2}+\sum\limits_{i=1}^4 n_{_i})}
{\Big[\prod\limits_{i=1}^4 n_{_i}!\Big]\Gamma(2-{D\over2}+n_{_1})\Gamma(2-{D\over2}+n_{_2})
\Gamma(2-{D\over2}+n_{_3})\Gamma({D\over2}+n_{_4})}\;.
\label{4loop-S21-14}
\end{eqnarray}

\item $J_{_{15}}=[6,7,8,9]$:
\begin{eqnarray}
&&\hspace{-1cm}\Phi_{_{[\tilde{1}345]}}^{(15)}=
y_{_4}^{{D\over2}-1}\sum\limits_{{\bf n}=0}^\infty
{c_{_{[\tilde{1}345]}}^{(15)}({\bf n})}f_{_{[\tilde{1}345]}}\;,\nonumber\\
&&\hspace{-1cm}c_{_{[\tilde{1}345]}}^{(15)}({\bf n})=
\frac{\Gamma(3-{D}+\sum\limits_{i=1}^4 n_{_i}) \Gamma(4-{3D\over2}+\sum\limits_{i=1}^4 n_{_i})}
{\Big[\prod\limits_{i=1}^4 n_{_i}!\Big]\Gamma({D\over2}+n_{_1})\Gamma(2-{D\over2}+n_{_2})
\Gamma(2-{D\over2}+n_{_3})\Gamma(2-{D\over2}+n_{_4})}\;.
\label{4loop-S21-15}
\end{eqnarray}

\item $J_{_{16}}=[7,8,9,10]$:
\begin{eqnarray}
&&\hspace{-1.6cm}\Phi_{_{[\tilde{1}345]}}^{(16)}=
\sum\limits_{{\bf n}=0}^\infty
{c_{_{[\tilde{1}345]}}^{(16)}({\bf n})}f_{_{[\tilde{1}345]}}\;,\nonumber\\
&&\hspace{-1.6cm}c_{_{[\tilde{1}345]}}^{(16)}({\bf n})=
\frac{\Gamma(4-{3D\over2}+\sum\limits_{i=1}^4 n_{_i}) \Gamma(5-{2D}+\sum\limits_{i=1}^4 n_{_i})}
{\prod\limits_{i=1}^4 n_{_i}!\Gamma(2-{D\over2}+n_{_i})}\;.
\label{4loop-S21-16}
\end{eqnarray}

\end{itemize}

\end{document}